\documentclass[review]{elsarticle}

\makeatletter
\let\c@author\relax
\makeatother

\expandafter\let\csname ver@natbib.sty\endcsname\relax

\usepackage{lineno,hyperref}
\modulolinenumbers[5]

\usepackage{graphicx}% Include figure files
\usepackage{dcolumn}% Align table columns on decimal point
\usepackage{bm}% bold math
\usepackage{multirow}
\usepackage{textcomp}
\usepackage[]{subfigure}

\usepackage{booktabs}
\usepackage{array}
\usepackage{pgf}
\usepackage{graphicx}
\usepackage[absolute,overlay]{textpos}
\usepackage{calc}
\usepackage{amsmath}
\usepackage{amsfonts}
\usepackage{amssymb}
\usepackage{textcomp}
\usepackage{tikz}
\usepackage{gensymb}

\usepackage{color}
\definecolor{Sun}{rgb}{0.164,0.126,0.322}
\definecolor{Green}{rgb}{0,0.300,0.300}
\definecolor{Red}{rgb}{0.4,0,0}
\definecolor{Grey}{RGB}{105,105,105}
\definecolor{Grey215}{RGB}{215,215,215}
\definecolor{White}{rgb}{1,1,1}

\usepackage{soul}

\usepackage{csquotes}

\newcommand{\commentCWS}[1]%
{\textsf{\textcolor{blue}{#1$^{\mathrm{CWS}}$}}}
\newcommand{\commentAV}[1]%
{\textsf{\textcolor{red}{#1$^{\mathrm{AV}}$}}}
\newcommand{\correction}[1]%
{\textsf{\textcolor{red}{#1}}}

\begin{document}

\begin{frontmatter}

\title{Atomistic Insights Into Cluster Strengthening in Aluminum Alloys}

\author[monash,ubc]{A.~de Vaucorbeil\corref{cor1}}
 \ead{alban.devaucorbeil@monash.edu}
\author[ubc]{C. W. Sinclair}
\author[ubc]{W. J. Poole}

\cortext[cor1]{Corresponding Author}

\address[monash]{Department of Materials Science and Engineering, Monash University, Clayton 3800, VIC, Australia}
\address[ubc]{Department of Materials Engineering The University of British Columbia, 309-6350 Stores Road, Vancouver, Canada} 

\begin{abstract}
In certain naturally aged aluminum alloys, significant strengthening can be obtained due to the decomposition of a super-saturated solid solution into clusters. The origins of such strengthening remain unclear due to the challenge of differentiating solute cluster strengthening from solid solution or precipitate strengthening.
To shed light on the origin of cluster strengthening in aluminum alloys, the interaction between the smallest possible type of clusters (\emph{i.e.} dimers) and moving dislocations in a model Al-Mg alloy is studied using atomistic simulations. Additionally, theoretical models for both the parelastic and dielastic interactions between clusters and dislocations is used to identify which factor among order strengthening, elastic interaction, and change of stacking fault energy controls cluster strengthening. The comparison of the results from these models to that of the atomistic simulations show that in the case of Mg dimers, the strength of the strongest ones are dominated by the dielastic contribution through the change of stacking fault energy.

\end{abstract}


\end{frontmatter}

\date{\today}
%-----------------------------------------------------------------------------------------------------------------------------------------------------------------------------

\section{Introduction}

Low temperature aging of some aluminum alloys leads to significant strengthening due to the decomposition of a super-saturated solid solution without the formation of a distinct second phase~\cite{RSP97,P95}.  This strengthening, commonly referred to as `cluster hardening', can contribute up to 70\% of the peak strength in some alloys~\cite{RSP97}. While atom probe tomography experiments have revealed that clustering in ternary alloys often involves the formation of co-clusters (e.g. Mg-Cu or Mg-Si)~\cite{RH00,MVRSP13,}, clustering (and cluster hardening) has also been reported in binary alloys~\cite{SKT82,OO84,KN63}.   

Given the indistinct, atomistic nature of solute clusters and the challenge of distinguishing cluster strengthening from solid solution and precipitation strengthening, the development of a fully physical model for cluster hardening has remained elusive. Measurements have revealed (e.g. \cite{MSFDR10}) that in some alloys the clusters contributing most significantly to strengthening contain only 2-4 solute atoms.  This has led several groups  to consider solute dimers (two solute elements as nearest neighbours) as the prototype of clusters for strengthening model development (e.g. \cite{SCR12,ZHLSL15,Z14}). The development of such a model is challenged by the inherent need for atomistic information to make it fully quantitative (e.g. \cite{Z14}) and also by the necessity for such atomistic data to be scaled up in order for the model to be useful for the prediction of the macroscopic yield/flow stress.

Several phenomena can contribute to the glide resistance provided by  a solute cluster.  Starink et al.~\cite{SW09,SCR12} considered that the strength of co-clusters could arise from a combination of modulus strengthening and order strengthening.  It was argued that modulus strengthening should only account for $\approx$10\% of the cluster strength in the case of Mg-Si and Mg-Cu co-clusters and therefore that the so-called order strengthening should dominate cluster strength.  Order strengthening, due to the change in atomic topology associated with a cluster straddling the glide plane, should be dominated by the enthalpy of cluster formation.  This model has been recently re-visited by Zhao~\cite{Z14} who noted that the elastic interaction between a dislocation and solute cluster, ignored in the model developed in~\cite{SW09}, should also provide a large contribution to the glide resistance.

Such phenomenological models require inherently atomistic information in order for them to be fully quantitative.  This requires either direct atomistic simulations or simplifying estimates in order for them to be quantitative.  Fortunately, atomic scale computer simulations of solute-dislocation interaction now allow to test the effects of atomic scale topology and energetics using simpler models (e.g. molecular dynamics/statics~\cite{PRBM06,PP11}, and/or Monte Carlo~\cite{XP06} based on semi-empirical potentials~\cite{DRS14}). 

Preliminary atomistic studies via molecular statics simulations have focused on the interaction of edge dislocations and solute dimers in binary Ni-Al~\cite{PP11} and Al-Mg~\cite{P09} alloys.  In  these studies the effect of the orientation of the dimer (Al-Al~\cite{PP11} or Mg-Mg~\cite{P09}) with respect to the glide plane and dislocation line was analyzed.  In both cases the results suggest that the change in topology of the dimer following shearing by the dislocation contributes only a small amount to the glide resistance of the dislocation~\cite{PP11}.

The goal of this work is to use atomistic simulations to predict the glide resistance of solute dimers in a model Al-Mg alloy.  From these results, the elastic and `order' or `chemical' contributions to the strength are evaluated separately in an attempt to identify the critical factors controlling cluster strengthening in the Al-Mg system.  %From this insight suggestions are made regarding the way in which one might construct alloys to maximize cluster strengthening.  

\section{Methodology:}

%\subsection{Simulation Box, Boundary and Loading Conditions}

\begin{figure}
  \centering
	\includegraphics[width=0.8\textwidth]{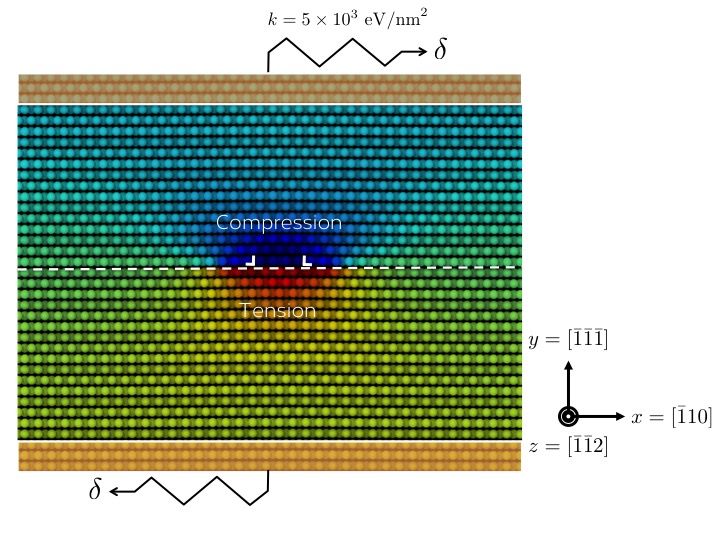}
	\caption{Truncated illustration of the simulation box used.  The colours schematically illustrate the hydrostatic stress field induced by the presence of the split edge dislocation. }
	\label{fig:schemdisl}
\end{figure}

Molecular statics simulations were performed on simulation boxes oriented as illustrated in figure \ref{fig:schemdisl}, the length of the box edges being $L_x=22~nm$, $L_y=28~nm$, and $L_z=6~nm$.  These dimensions were obtained as a compromise between computational speed and minimization of image stress effects, periodicity having been enforced in the $x$ and $z$ directions\cite{vaucorbeil15}.  Into this box an edge dislocation having a Burger's vector of $b = a/2[\bar{1}10]$ and line direction $[\bar{1}\bar{1}{2}]$ was inserted as described in~\cite{BOR09}.  The simulation box was subjected to energy minimization while the box size was modified so as to bring the macroscopic pressure on the simulation box to less than 10 MPa.  Upon energy minimization, the dislocation split into two Shockley partials dislocations separated by a stacking fault.  While only edge dislocations were studied here, it has been shown by Patinet and Proville~\cite{PP11} that the solute-dislocation interactions are similar for both edge and screw dislocations. This is explained by the fact that in aluminium alloys, both screw and edge dislocations split into partial dislocations of mixed character. These partial dislocations have the same fraction of edge and screw components regardless of the character of the perfect dislocation from which they arise.  As it will be shown in the following, the strongest interactions occur here when solutes are close to the cores of the partial dislocations. It is thus the solute-partial dislocation interaction that is more important.  However, if longer-ranged interactions were dominant, the net character of the dislocation (edge or screw) would be more important as the sums of the stress fields of the partial dislocations (felt for solutes located far from one of the partials) will be different in the case of edge and screw dislocations.

In order to directly measure the energy-distance and force-distance profiles corresponding to dislocation/dimer interactions, a mixed boundary condition was used to impose deformation on the simulation box \cite{Rodney07}.  The centre of mass of the top and bottom two atomic layers were attached to a linear spring having a stiffness of $k=5\times 10^3~eV/nm^2$ and an incremental displacement of 7 $\times 10^{-3} nm$ was imposed in the $x$ direction before each step of minimization.    As described in \cite{Rodney07}  this loading condition allows atoms in the upper and lower surfaces to adapt to the plastic strain produced during dislocation glide. 

An embedded atom method interatomic (EAM) potential for Al-Mg developed by Liu and Adams~\cite{LA98} was used here. It has a cutoff distance of 6.668~$\AA$ and has been widely used before on the study of solid solution strengthening in Al-Mg alloys \cite{P09,PP11,OHC06,XP06,ZP04}.  The predicted elastic constants and stacking fault energy for pure Al from this potential are given in Table \ref{table:elastic_constants} while the temperature and composition dependence (for up to Al-10at\%Mg) of the elastic constants and stacking fault energy predicted by this potential have been reported by Dontsova \emph{et al.} \cite{dontsova14}.  All simulations were performed using LAMMPS \cite{lammps_main} while visualization and dislocation position were extracted using Ovito \cite{ovito}.

\begin{table}
 \caption{Single crystal elastic constants and intrinsic stacking fault energy ($E_{SF}$) of pure aluminum from experiments and predicted by the EAM potential used in this study~\cite{LA98} }.
\resizebox{\textwidth}{!}{%
\begin{tabular}{c|cccc}
\hline
& $C_{11}$ & $C_{12}$ & $C_{44}$ & $E_{SFE}$ \\
\hline
Experimental (4~K) ~\cite{VMSB64} & 116 GPa  & 64.8 GPa & 30.9 GPa & 0.120 - 0.144 J/m$^2$ ~\cite{WP71,R82} \\
EAM Potential (0~K) ~\cite{LA98} & 119 GPa & 62.3 GPa & 34.9 GPa & 0.135 J/m$^2$ \\
\hline
  \end{tabular}}
    \label{table:elastic_constants}
\end{table}

For the purposes of this work, Mg atoms were introduced into the simulation domain shown in Figure \ref{fig:schemdisl} as `dimers' being defined by two solute Mg atoms occupying either first nearest neighbour positions along $\left<110\right>$ directions or second nearest neighbour positions along $\left<100\right>$ directions.  The extension of these results to trimers is reported elsewhere \cite{vaucorbeil15}.  %To look at only the two dimers with the highest strength, one of the Mg atoms comprising the dimer was located on the glide plane.  The second atom then either sat on the same plane, one plane above or one plane below.  As has been shown elsewhere, the strength of dimers drops rapidly as the centre of mass of the dimer moves away from the glide plane \cite{vaucorbeil15}.

\section{Summary of Dimer Strengths}

\begin{figure}[htbp]
\centering     %%% not \center
\subfigure[]{\label{fig:dimers_a}\includegraphics[width=1\textwidth]{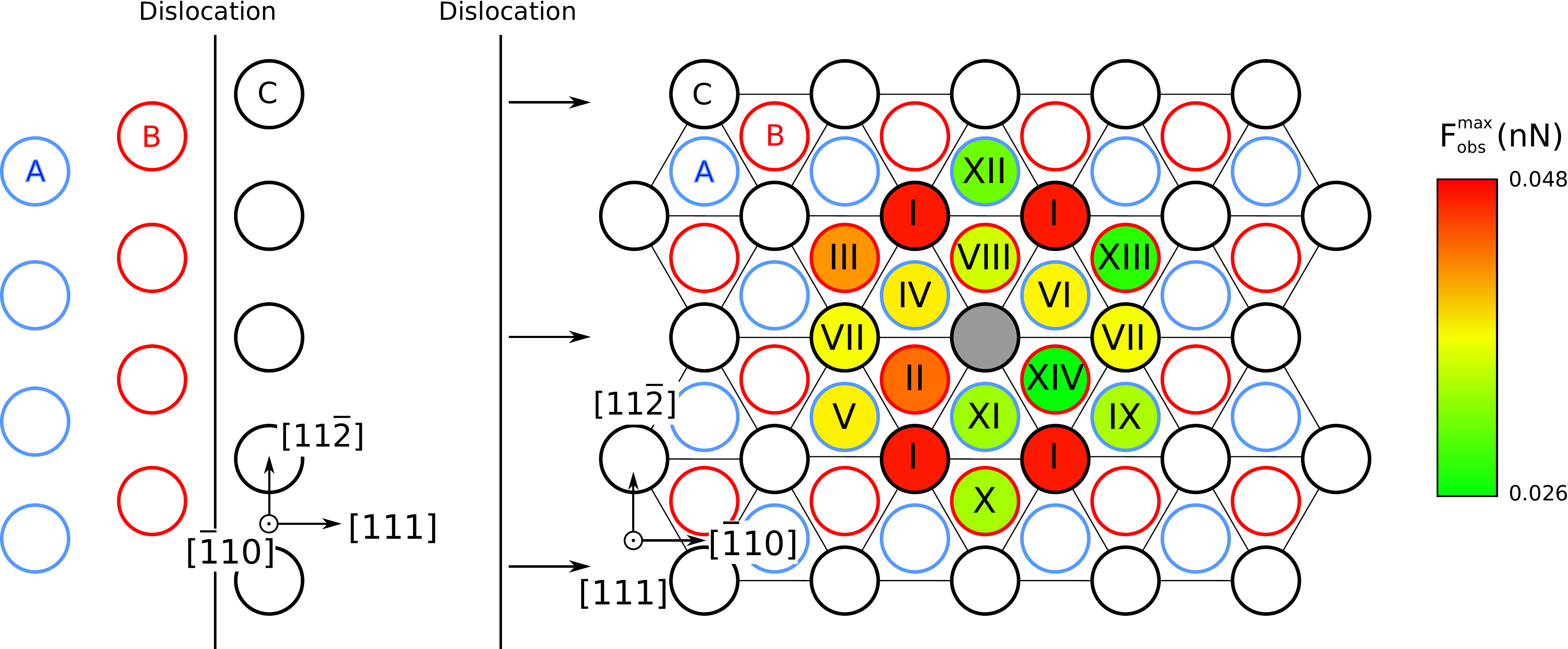}}
\subfigure[]{\label{fig:dimers_b}\includegraphics[width=0.8\textwidth]{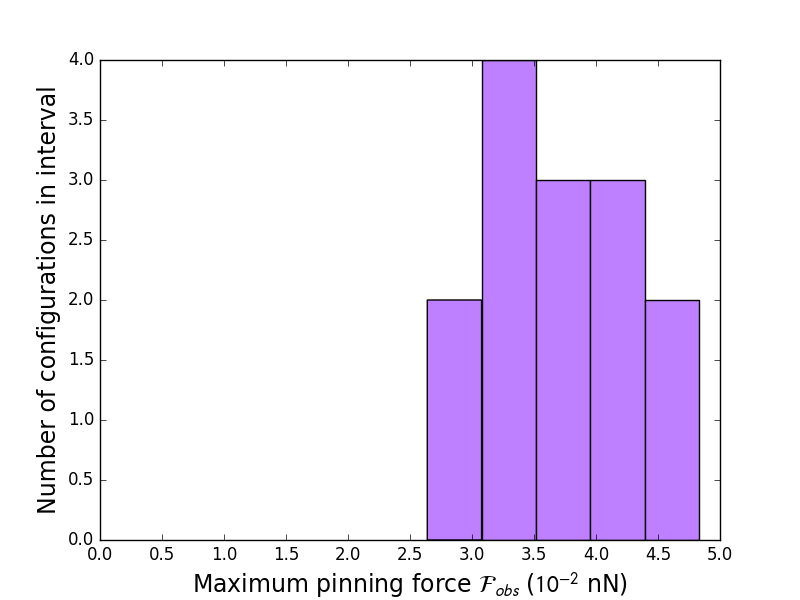}}
\caption{a) Schematic illustration of the glide plane showing three atomic planes; the black circles indicate atoms lying on the $\{111\}$ glide plane, the red circles indicate a $\{111\}$ plane one plane above the glide plane and the blue circles indicate atoms located one plane below the glide plane.  The grey central atomic position shows one of the positions of the Mg atoms comprising the dimers studied.  The second atom in the dimer was taken to be located at the atomic positions indicated by Roman numerals.  The filled colours at these sites indicates the strength of the dimer to shearing by a dislocation coming from the left hand side.  b) The distribution of strengths of the dimers studied factoring in the multiplicity of non-unique dimer locations.}
\label{fig:dimers}
\end{figure}

Figure \ref{fig:dimers_a} illustrates the strength of the 14 different configurations of dimers that were investigated as well as their detailed positioning with respect to the glide plane. This represents all possible combinations where the dimers are nereast neighbours. Each dimer is composed of two Mg atoms, one located at in the central (grey) position, the second in one of the atomic positions marked by the Roman numerals.
For each on these configurations, and according to their relative positions, Mg atoms were introduced in a simulation box already containing a split edge dislocation as far as possible from the location of the dislocation (\textit{i.e.} at around $11~nm$ from the center of the dislocation in the $[\bar{1}10]$ direction). Then, an energy minimization was performed to remove any change in the macroscopic pressure caused by the addition of the Mg atoms, and allow local rearrangment of atoms around the dimer. Finally, the box was loaded according to the method described above, causing the dislocation to glide and interact with the dimer, and the energy and force profiles recorded during the dislocation glide.
On Figure \ref{fig:dimers_a}, the colour at each of the sites labeled by a Roman numeral indicates the corresponding strength of that dimer reported here as the pinning force obtained from the maximum measured macroscopic virial stress ($\tau_{xy}$) as,

\begin{equation}
F_{dimer} = \left(\tau_{xy} - \tau_p\right)bL_z
\end{equation}

\noindent where $\tau_p$ is the Peierls stress, measured here to be 2.8 MPa from the simulations without any Mg atoms.  The distribution of all of these strengths weighted by the total number of possible configurations for each dimer is given in Figure \ref{fig:dimers_b}

The distribution of strengths shown in Fig. \ref{fig:dimers_b} can be compared with the strength to overcome  a single solute atom, located at the position indicated by the grey atom in Fig. \ref{fig:dimers_a} which was found to be 0.026 nN.   Thus, the two strongest dimers appear to be approximately twice as strong as a single solute atom at the same location.  This result is similar to that of Patinet and Proville~\cite{PP11} who found that the maximum pinning force for the strongest dimer (dimer I here) was $0.047~nN$~\cite{PP11} ($0.048~nN$ in the current work). 

\begin{figure}
  \centering
  \subfigure[Before passage of the dislocation.]{\label{fig:dimer1_a}\includegraphics[width=0.48\textwidth]{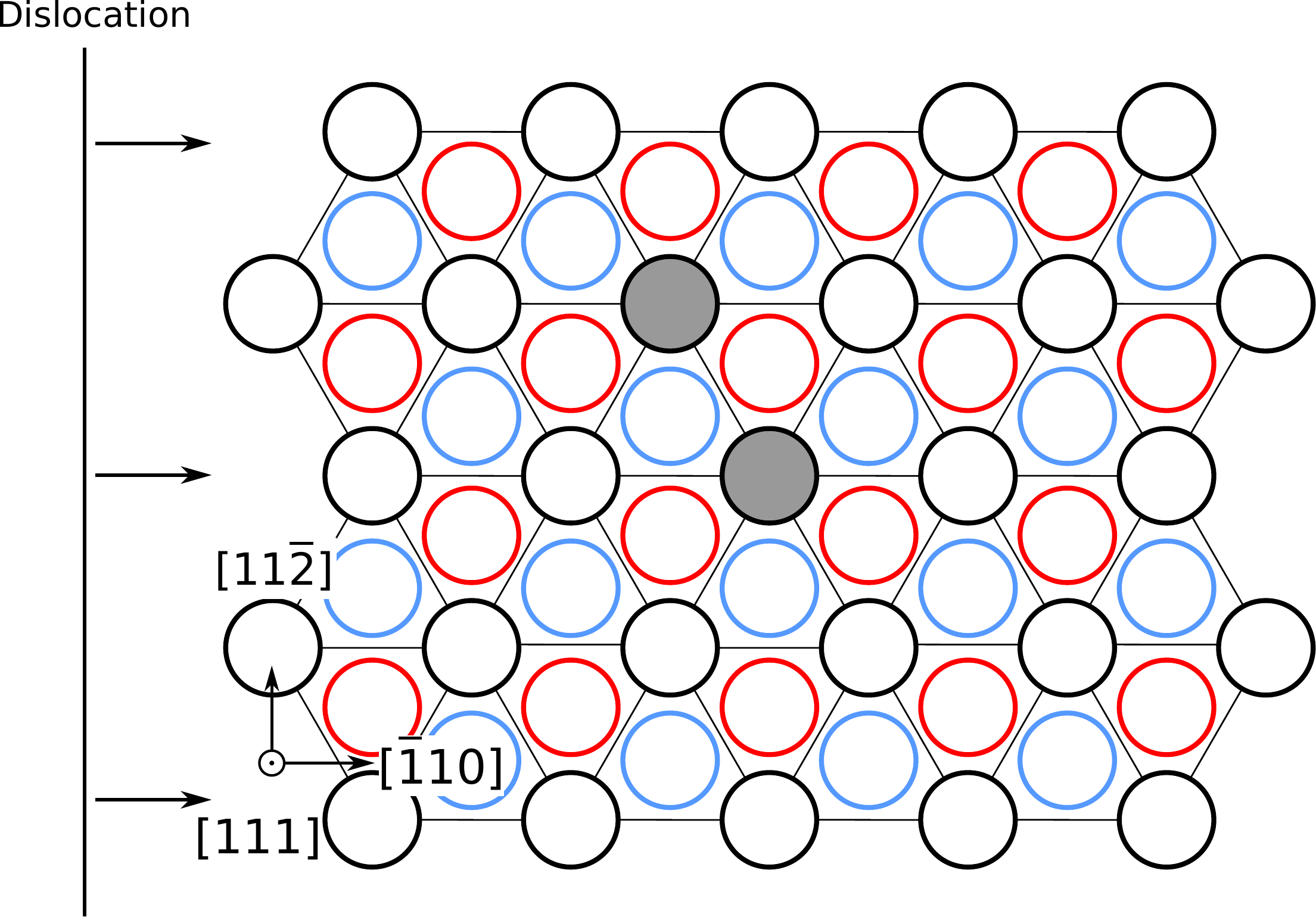}}
  \subfigure[After the dislocation has passed.]{\label{fig:dimer1_b}\includegraphics[width=0.48\textwidth]{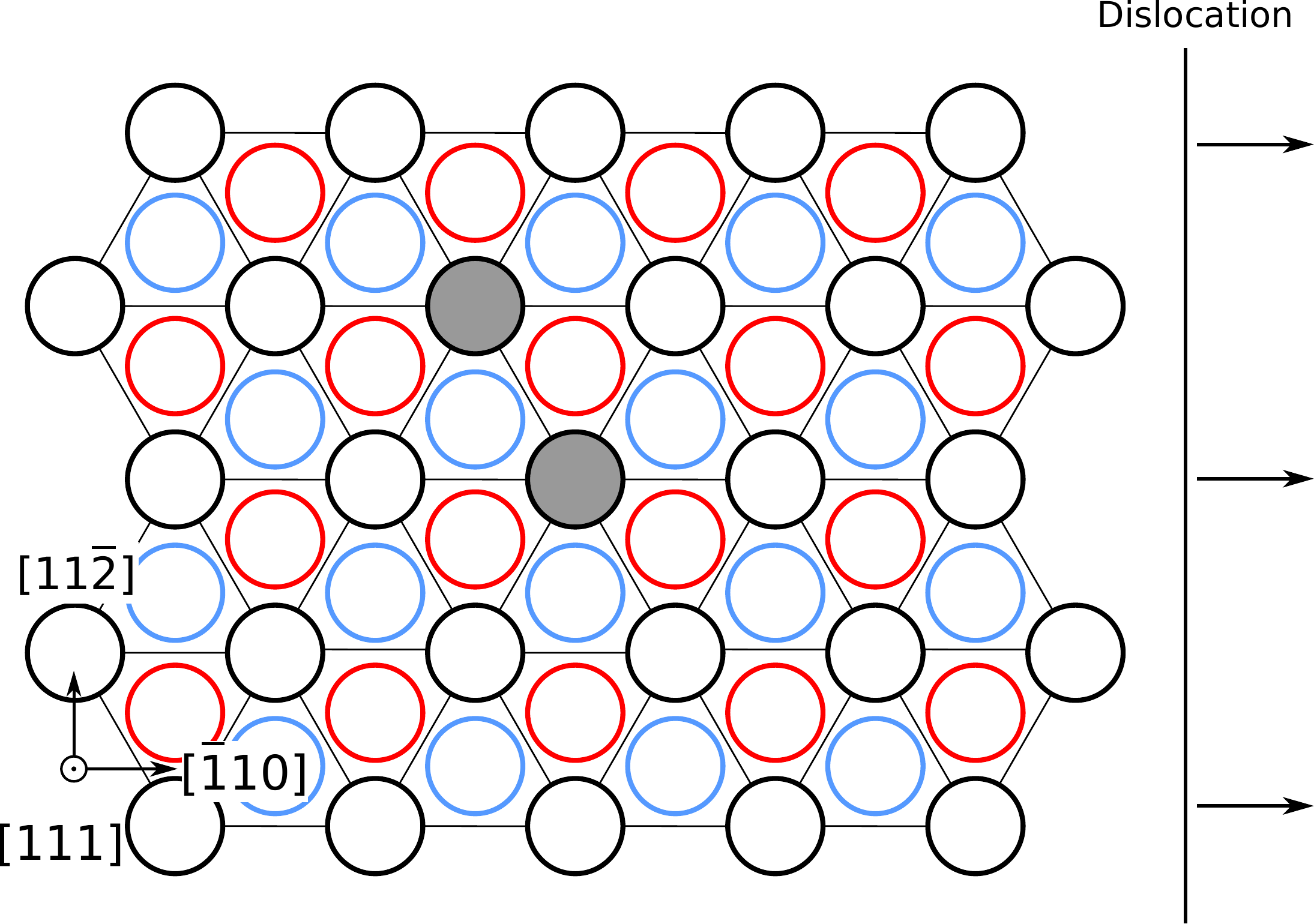}}
	\caption{Illustration of the change in position and configuration of Mg atoms in dimer I after passage of an edge dislocation as viewed looking down on the glide plane. }
	\label{fig:dimer1}
\end{figure}

The two strongest dimers, dimer I and dimer II in Fig. \ref{fig:dimers_a}  have quite different configurations with respect to the passage of a dislocation.   The strongest, dimer I, has both Mg atoms located on the same glide plane above the plane of the dislocation in the region of compressive hydrostatic stress.  The separation between these two atoms does not, therefore, change after the passage of the dislocation (see Fig. \ref{fig:dimer1}) meaning that the concept of order strengthening as applied in the literature~\cite{SW09} should not apply.  

\begin{figure}
  \centering
  \subfigure[Before passage of the dislocation.]{\label{fig:dimer2_a}\includegraphics[width=0.48\textwidth]{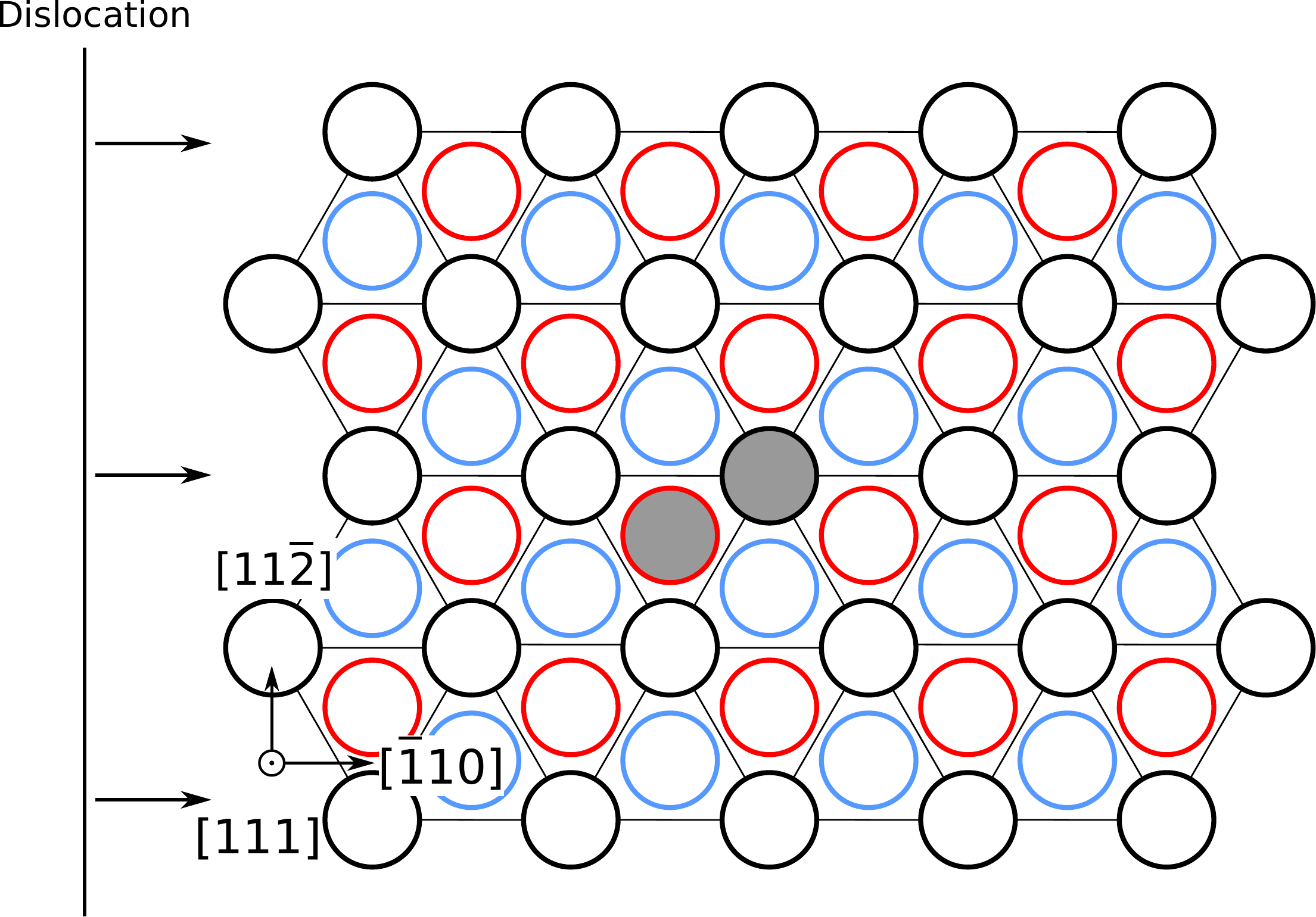}}
  \subfigure[After the dislocation has passed.]{\label{fig:dimer2_b}\includegraphics[width=0.48\textwidth]{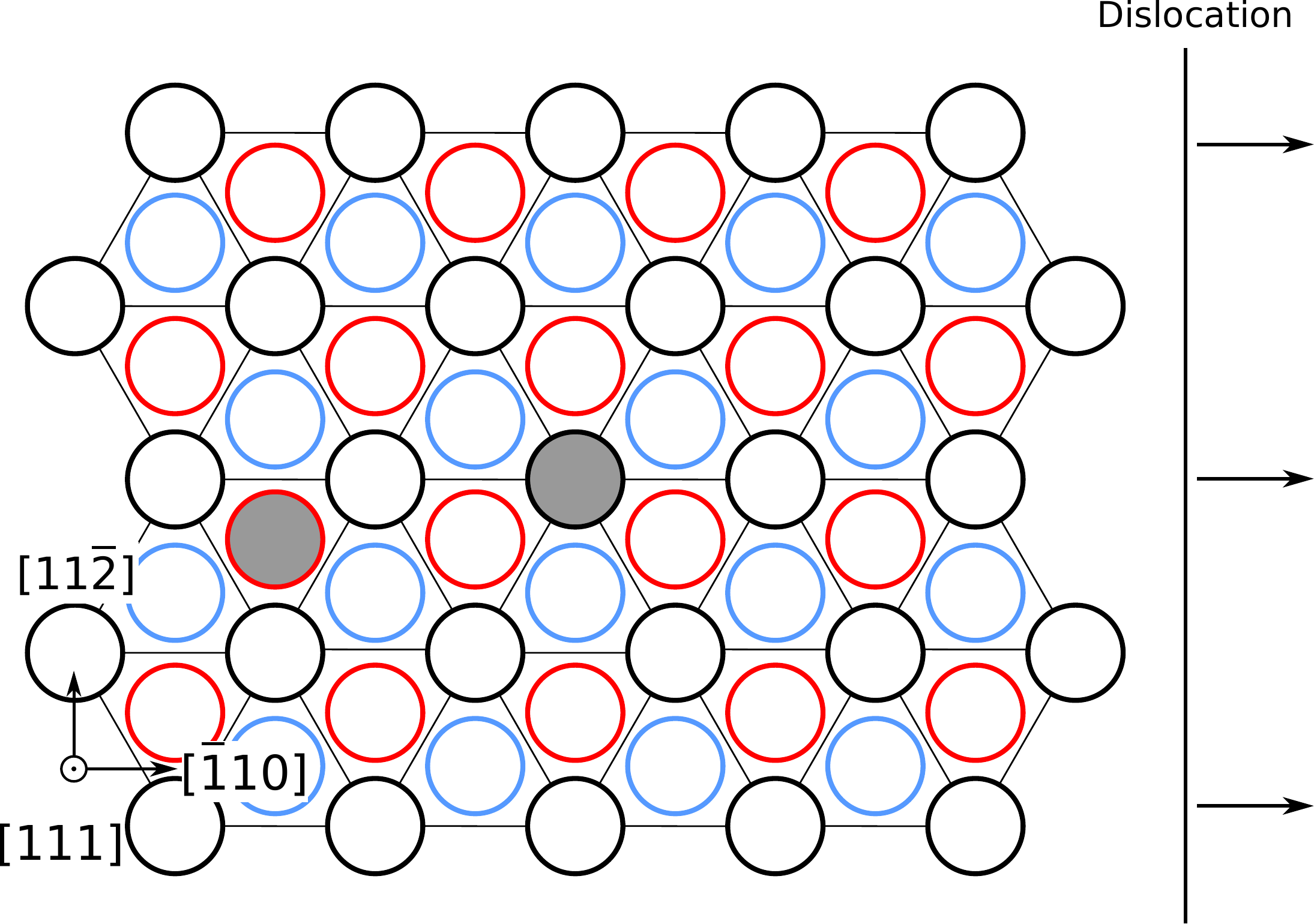}}
	\caption{Illustration of the change in position and configuration of Mg atoms in dimer II after passage of an edge dislocation as viewed looking down on the glide plane. }
	\label{fig:dimer2}
\end{figure}

In contrast, dimer II is composed of two Mg atoms that straddle the glide plane, this leading to their separation based on the passage of the dislocation as shown in Fig. \ref{fig:dimer2}.  In this case, one may expect an elastic interaction between the solute and the dislocation as well as the `order' strengthening contribution proposed by Starkink \emph{et al.}~\cite{SW09,SCR12}   arising from the change in atomic configuration.

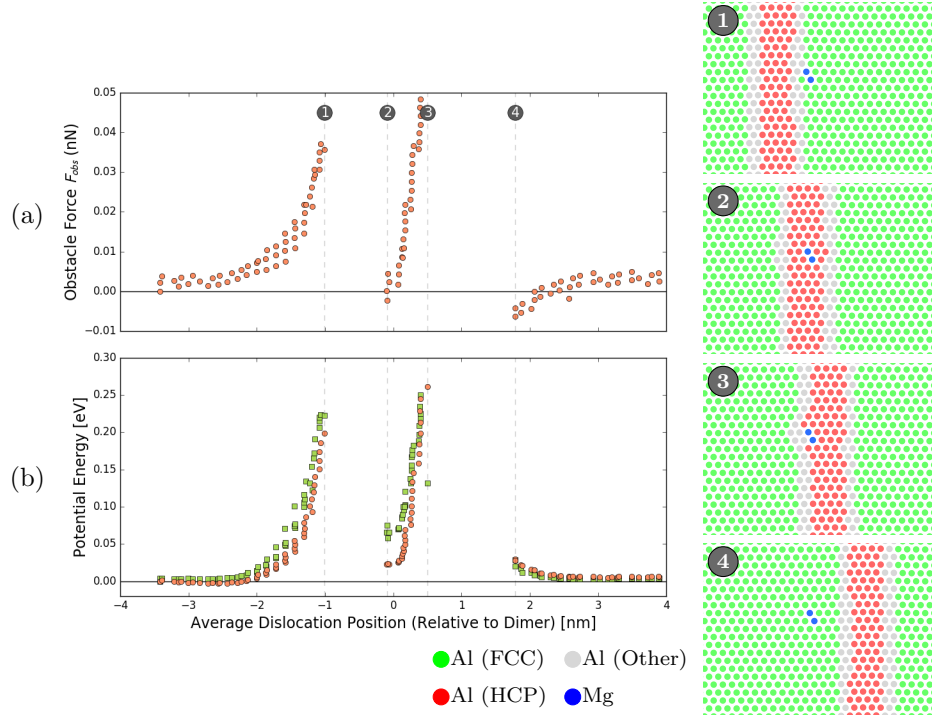
\begin{figure}[htbp]
\centering     %%% not \center
\begin{tikzpicture}
\node at (0,0) {\includegraphics[width=0.7\textwidth]{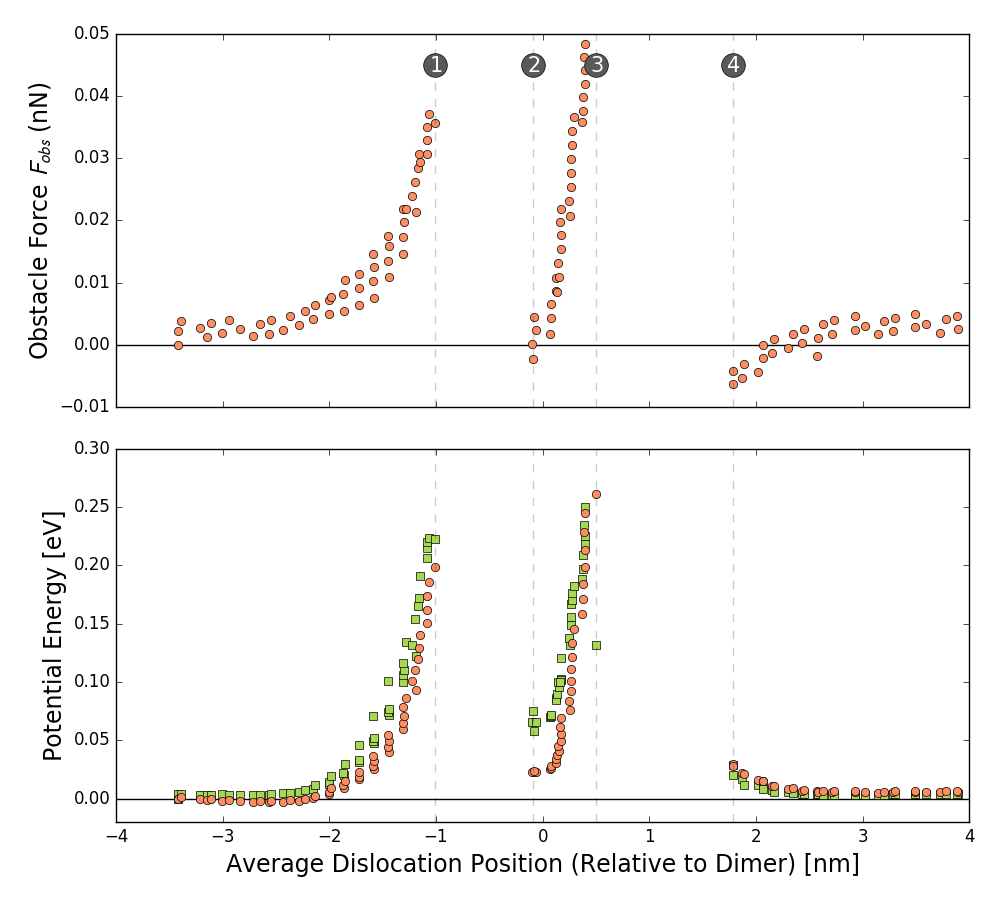}};
\node at (-4.5,1.9) {(a)};
\node at (-4.5,-1.6) {(b)};
\node at (6,3.6) {\includegraphics[width=0.25\textwidth]{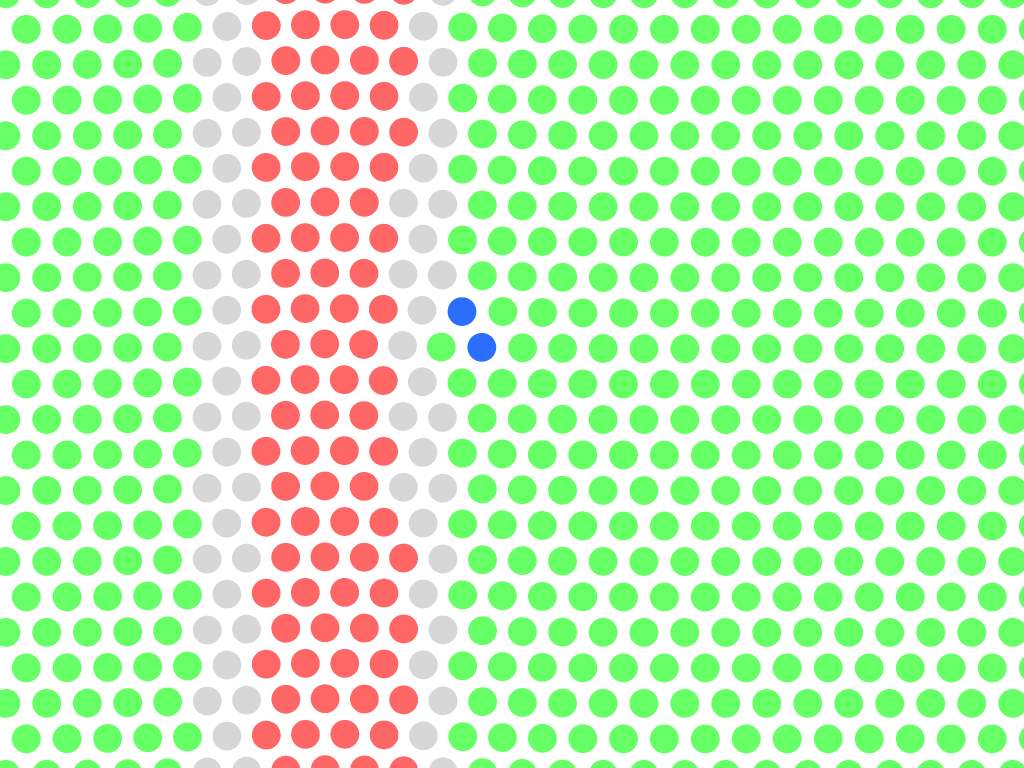}};
\node at (6,1.2) {\includegraphics[width=0.25\textwidth]{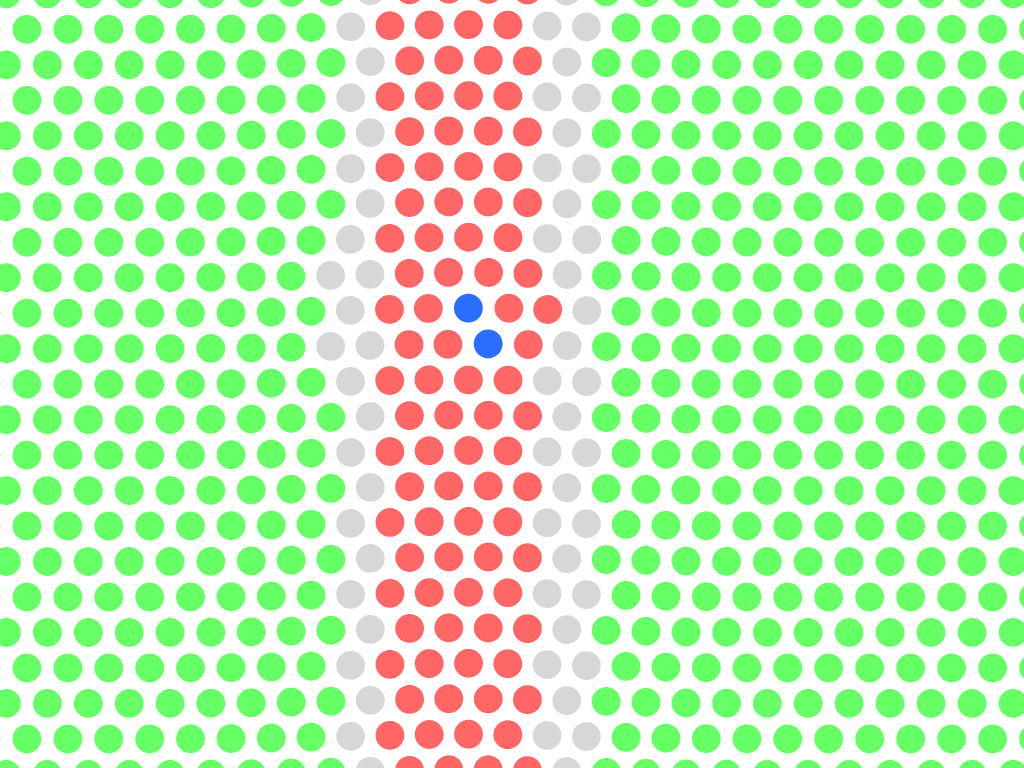}};
\node at (6,-1.2) {\includegraphics[width=0.25\textwidth]{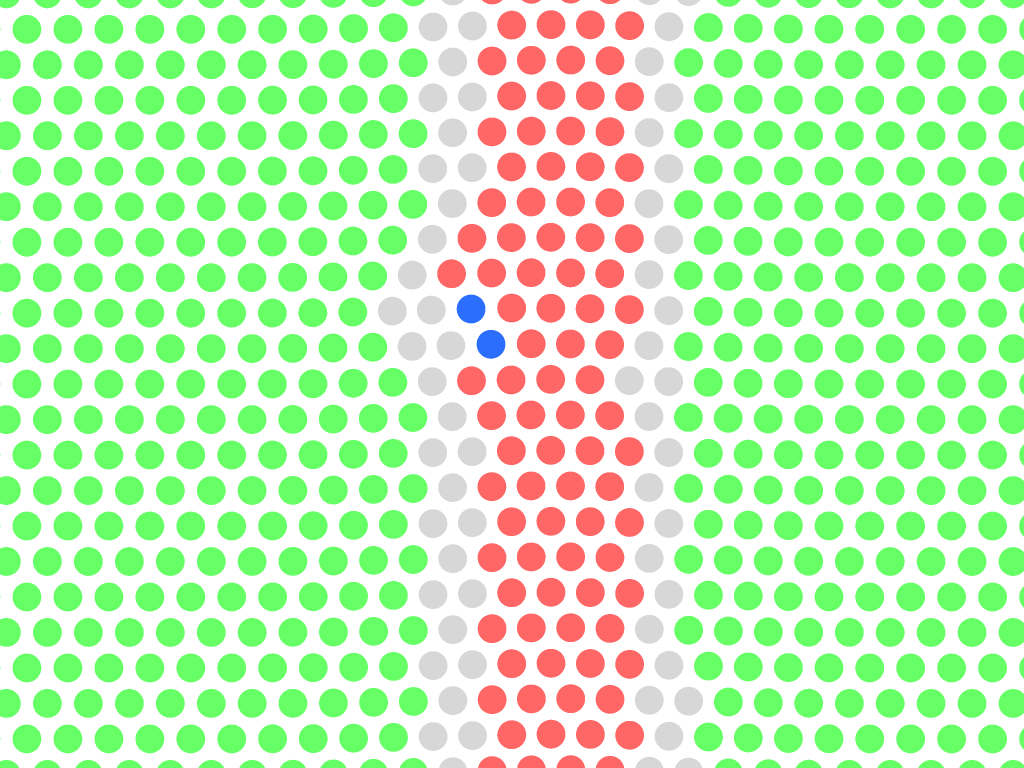}};
\node at (6,-3.6) {\includegraphics[width=0.25\textwidth]{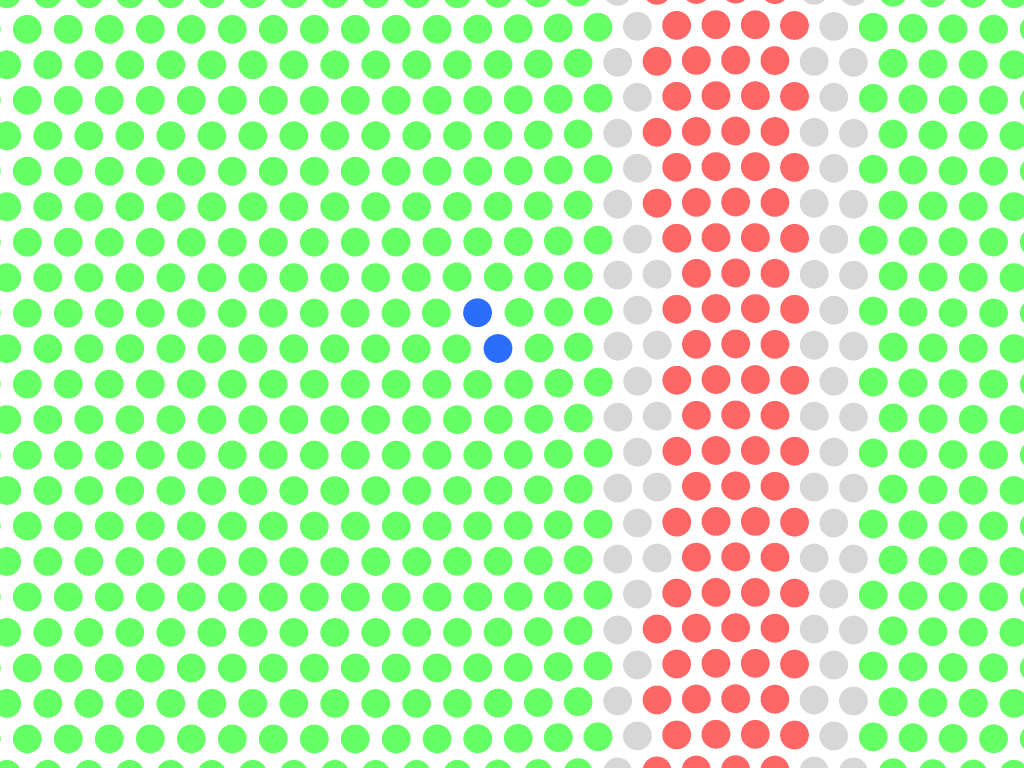}};
\node [shape=circle,draw,inner sep=1.5pt,fill=Grey] at (4.75,4.5) {\footnotesize\textcolor{white}{\textbf{1}}};
\node [shape=circle,draw,inner sep=1.5pt,fill=Grey] at (4.75,2.1) {\footnotesize\textcolor{white}{\textbf{2}}};
\node [shape=circle,draw,inner sep=1.5pt,fill=Grey] at (4.75,-0.3) {\footnotesize\textcolor{white}{\textbf{3}}};
\node [shape=circle,draw,inner sep=1.5pt,fill=Grey] at (4.75,-2.7) {\footnotesize\textcolor{white}{\textbf{4}}};
\draw [green,fill=green] (1,-4) circle (3pt);
\node [right] at (1,-4) {\footnotesize{Al (FCC)}};
\draw [red,fill=red] (1,-4.5) circle (3pt);
\node [right] at (1,-4.5) {\footnotesize{Al (HCP)}};
\draw [Grey215,fill=Grey215] (2.75,-4) circle (3pt);
\node [right] at (2.75,-4) {\footnotesize{Al (Other)}};
\draw [blue,fill=blue] (2.75,-4.5) circle (3pt);
\node [right] at (2.75,-4.5) {\footnotesize{Mg}};
\end{tikzpicture}
\caption{Force distance and energy distance profiles for dislocation/dimer I interactions based on imposed shearing of simulation box (round symbols). The square symbols in the energy distance profile correspond to the addition of the elastic strain energy and the interaction energy obtained when the dimer is manually displaced from one atomic position to the next (see section \ref{sec:unraveling}).   In these plots, `0' on the x-axis represents the centre of mass of the dimer (prior to deformation) along the glide direction ($x$) direction.  The atomistic snapshots show the $\{111\}$ glide plane and illustrate the position of the dislocation, stacking fault and Mg atoms at the positions indicated by the dashed vertical lines in the force-distance and energy-distance plots.}
\label{fig:dimerprofile5}
\end{figure}

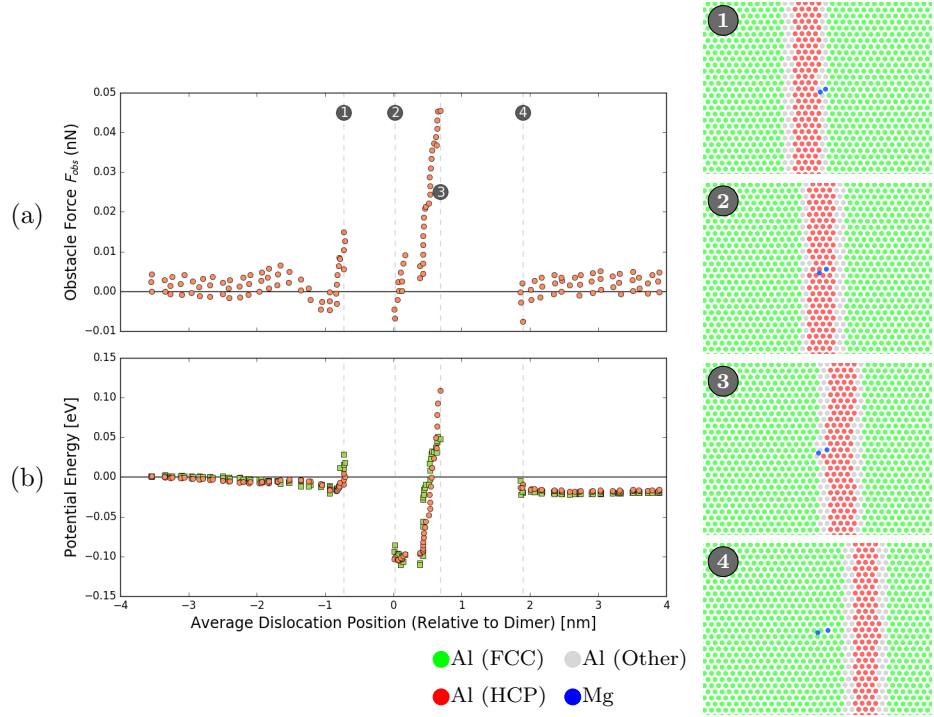
\begin{figure}[htbp]
\centering     %%% not \center
\begin{tikzpicture}
\node at (0,0) {\includegraphics[width=0.7\textwidth]{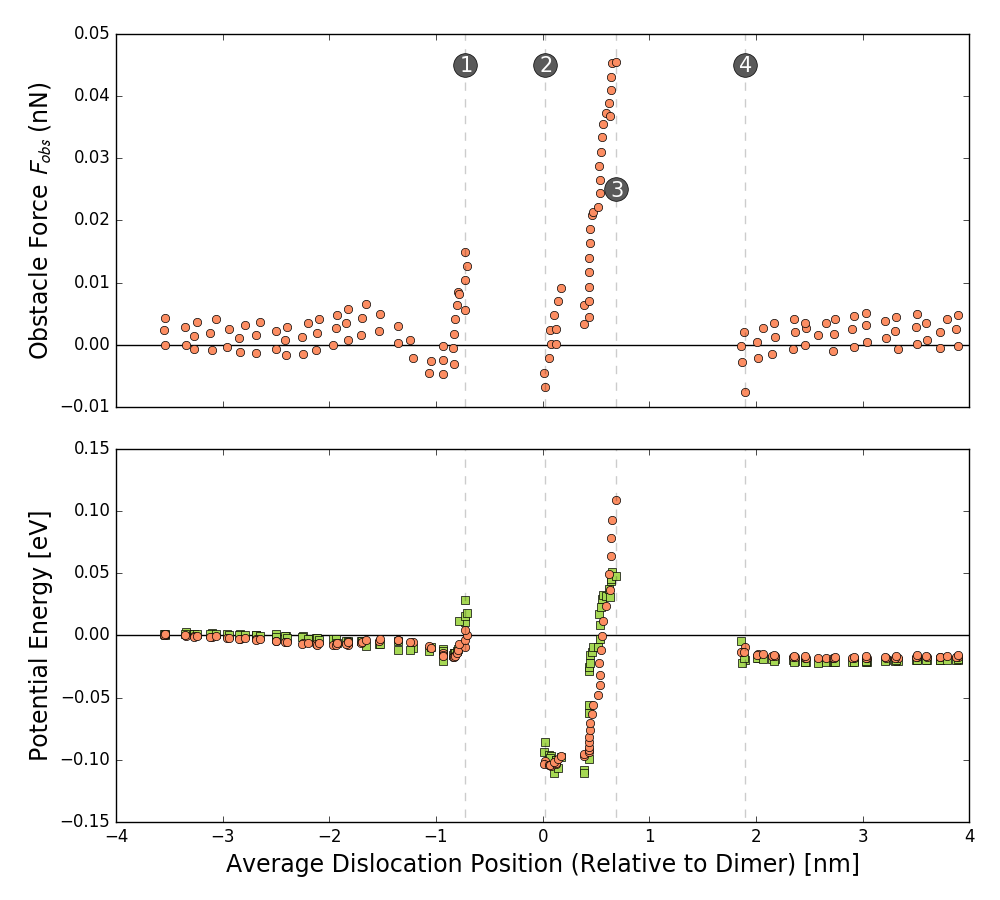}};
\node at (-4.5,1.9) {(a)};
\node at (-4.5,-1.6) {(b)};
\node at (6,3.6) {\includegraphics[width=0.25\textwidth]{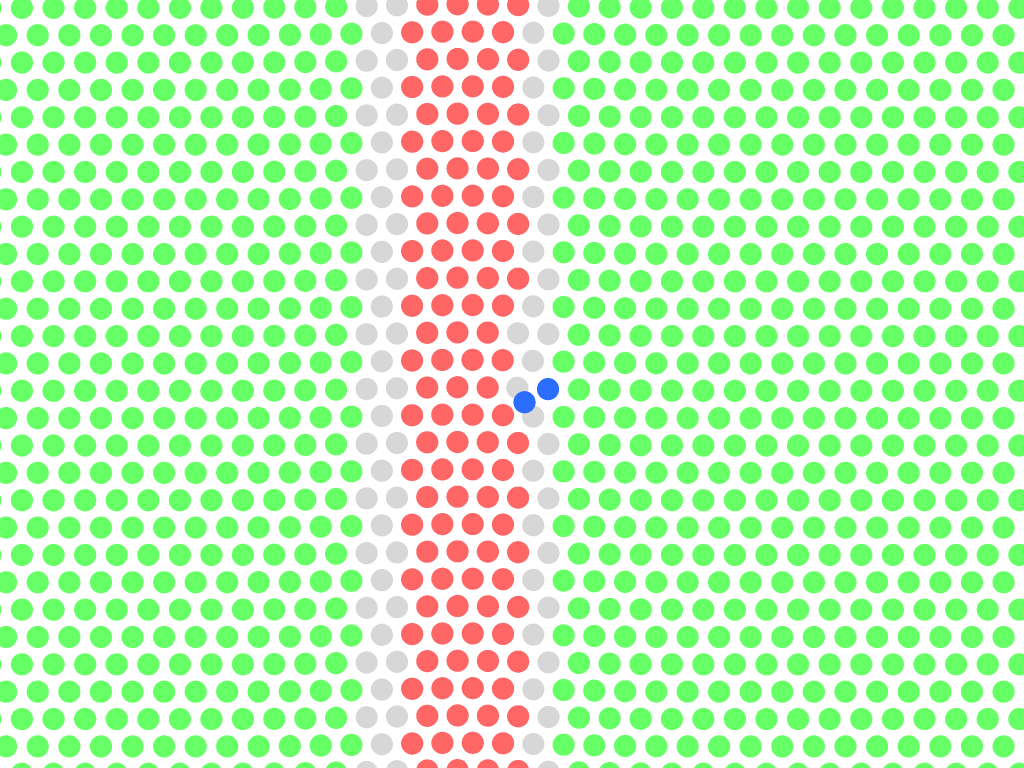}};
\node at (6,1.2) {\includegraphics[width=0.25\textwidth]{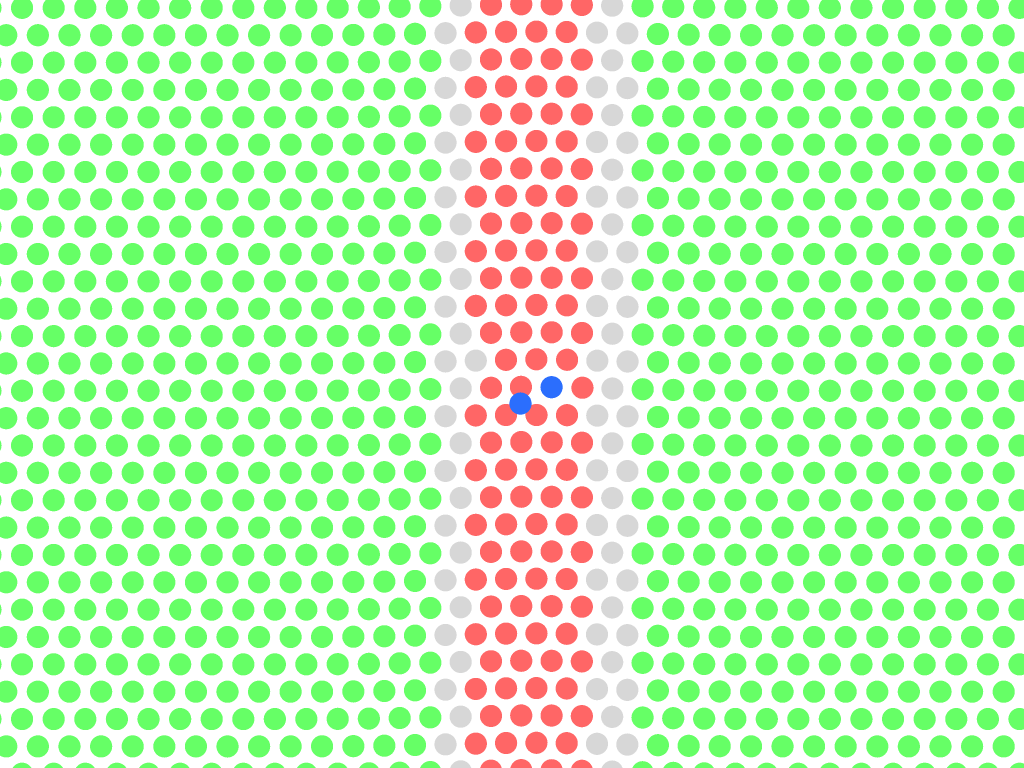}};
\node at (6,-1.2) {\includegraphics[width=0.25\textwidth]{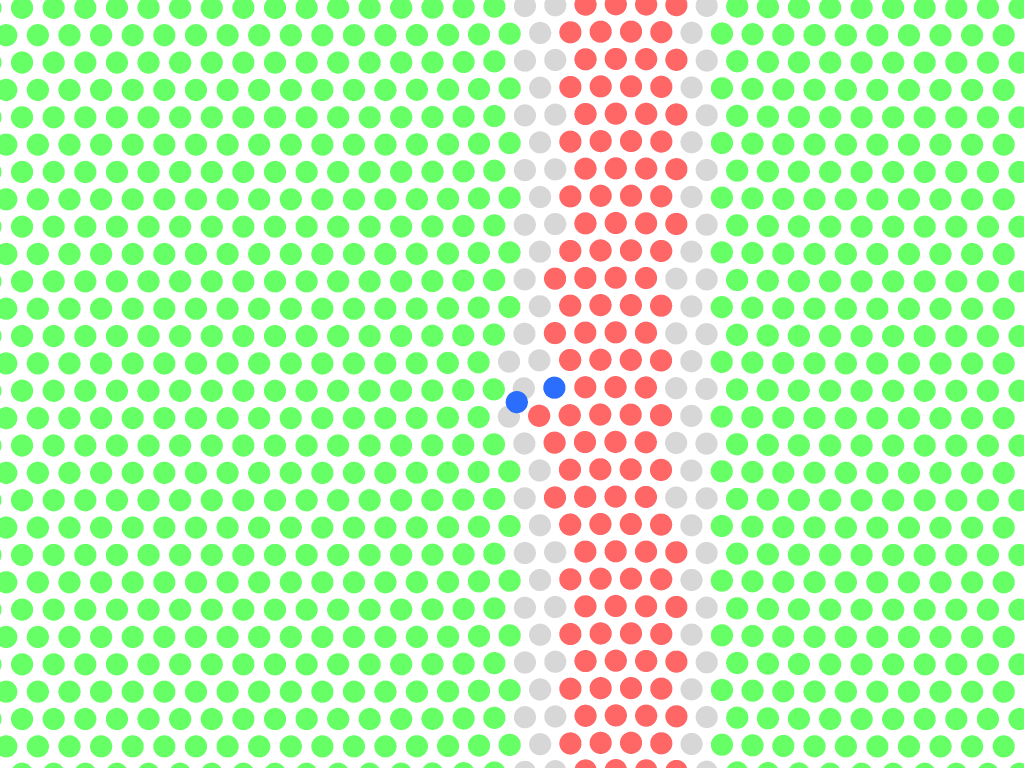}};
\node at (6,-3.6) {\includegraphics[width=0.25\textwidth]{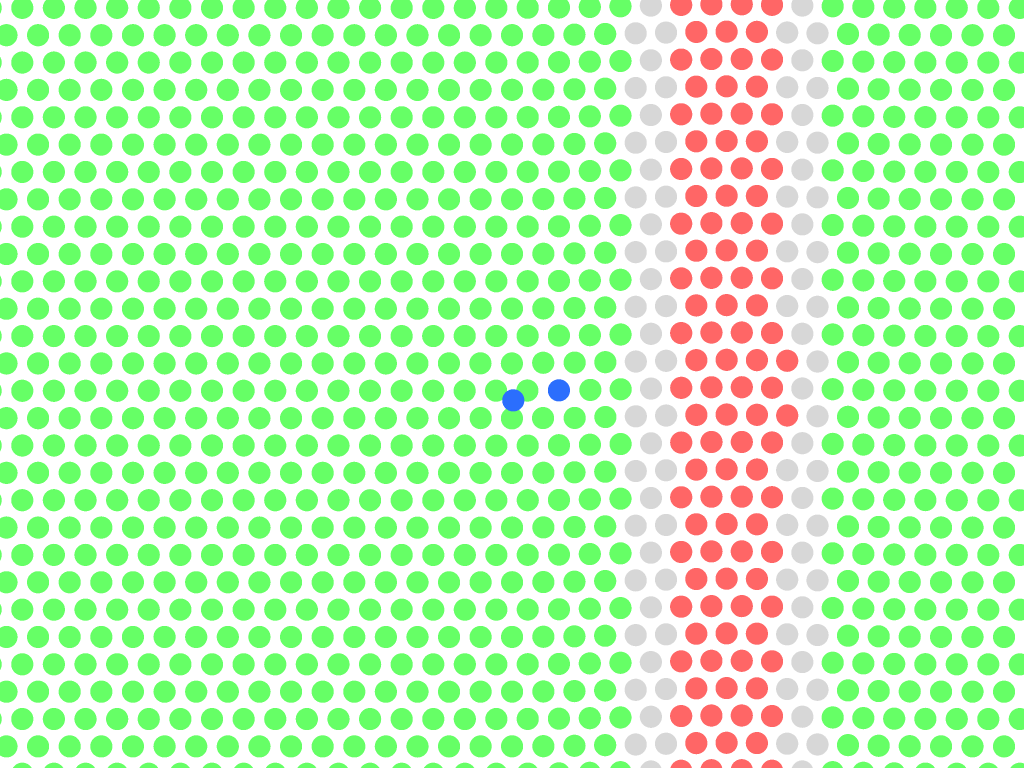}};
\node [shape=circle,draw,inner sep=1.5pt,fill=Grey] at (4.75,4.5) {\footnotesize\textcolor{white}{\textbf{1}}};
\node [shape=circle,draw,inner sep=1.5pt,fill=Grey] at (4.75,2.1) {\footnotesize\textcolor{white}{\textbf{2}}};
\node [shape=circle,draw,inner sep=1.5pt,fill=Grey] at (4.75,-0.3) {\footnotesize\textcolor{white}{\textbf{3}}};
\node [shape=circle,draw,inner sep=1.5pt,fill=Grey] at (4.75,-2.7) {\footnotesize\textcolor{white}{\textbf{4}}};
\draw [green,fill=green] (1,-4) circle (3pt);
\node [right] at (1,-4) {\footnotesize{Al (FCC)}};
\draw [red,fill=red] (1,-4.5) circle (3pt);
\node [right] at (1,-4.5) {\footnotesize{Al (HCP)}};
\draw [Grey215,fill=Grey215] (2.75,-4) circle (3pt);
\node [right] at (2.75,-4) {\footnotesize{Al (Other)}};
\draw [blue,fill=blue] (2.75,-4.5) circle (3pt);
\node [right] at (2.75,-4.5) {\footnotesize{Mg}};
\end{tikzpicture}
\caption{Force distance and energy distance profiles for dislocation/dimer II interactions based on imposed shearing of simulation box (round symbols). The square symbols in the energy distance profile correspond to the addition of the elastic strain energy and the interaction energy obtained when the dimer is manually displaced from one atomic position to the next (see section \ref{sec:unraveling}).   In these plots, `0' on the x-axis represents the centre of mass of the dimer (prior to deformation) along the glide direction ($x$) direction. The atomistic snapshots show the $\{111\}$ glide plane and illustrate the position of the dislocation, stacking fault and Mg atoms at the positions indicated by the dashed vertical lines in the force-distance and energy-distance plots.}
\label{fig:dimerprofile4}
\end{figure}

While the peak force to bypass these two dimers is very similar, the force-distance and energy-distance profiles defining the interaction between an edge dislocation and the two dimers is quite different (Fig. \ref{fig:dimerprofile5} and \ref{fig:dimerprofile4}).  In the case of dimer I (both solute on compressive side of glide plane), when viewed from left to right (dislocation approaching the dimer) the force is seen to rise as $\sim r^{-2}$ as one might expect on the basis of an interaction between a centre of dilation and the hydrostatic stress field of the dislocation \cite{ } (Fig. \ref{fig:dimerprofile5}).  The sudden drop in force between stages 1 and 2 corresponds to the leading partial overcoming the dimer, leading to the dimer being situated within the dislocations stacking fault.  With continued loading, the stress rises again until the second partial overcomes the dimer. As soon as this happens, the stored elastic energy between stages 1 and 2 is released all at once causing the dislocation to glide until equilibrium is reached again at stage 4, hence the discontinuity between stages 3 and 4.
Owing to the lack of change in the configuration between the two solute atoms after the dislocation has passed, there is no net change in system energy once the dislocation moves away. 

In the case of dimer II a rather different force-distance and energy distance profile are observed.  In this instance, as the leading partial approaches the dimer, there is little change in energy and only small changes in force. Indeed, as the solute atoms are located on either (tensile and compressive) sides of the dislocation, their net elastic interaction almost cancels out.  Once the leading partial overcomes the dimer, the force is seen to rise rapidly as the trailing partial is forced to bypass the dimer.  Note that in this example, as can be seen from Fig. \ref{fig:dimerprofile4}(b), the potential energy of the system is higher at the beginning of the simulation (prior to stage 1) compared to the end of the simulation (post stage 4). This drop in system energy is a result of the change in configuration of solute atoms comprising the dimer, and is contrary to the case of dimer I (Fig. \ref{fig:dimer2}). As the dislocation glides through dimer II, the Mg atoms go from being first nearest neighbours (prior to stage 1) to being second nearest neighbours (after stage 4) which is a more favourable position as attested by this energy drop. This observation is consistent with experimental observations~\cite{SKT82} and prior reports related to this interatomic potential~\cite{ZP04,dontsova14}.

\section{Unraveling the Contributions to Dimer Strength}
\label{sec:unraveling}

The results in Figs. \ref{fig:dimers_a}, \ref{fig:dimerprofile5} and \ref{fig:dimerprofile4} are most directly described as a consequence of the change in system energy with dislocation position arising from i) elastic interaction between the dislocation and dimer (parelastic contribution~\cite{H96}) ii) configurational changes between the Mg atoms arising from the passage of the dislocation (dielastic contribution ~\cite{H96}) and iii) macroscopic elastic loading of the sample.  Below, we will present models for contributions (i) and (ii) with the aim of identifying the separate contributions attributable to each.  Unlike the calculations performed above, these models predict equilibrium energies of a system containing a dimer and a split edge dislocation as a function of their separation \emph{without considering any externally imposed force}.  Thus, these models are not directly comparable to the energy-distance curves shown in Figs.  \ref{fig:dimerprofile5} and \ref{fig:dimerprofile4} where the system is under the influence of an additional external applied force.   To bridge the gap between the direct calculations shown in Figs. \ref{fig:dimerprofile5} and \ref{fig:dimerprofile4} and the models developed below, a second set of fully atomistic calculations have been performed.  In these calculations, the same simulation box used to produce the results in Figs. \ref{fig:dimerprofile5} and \ref{fig:dimerprofile4} was employed. However, rather than moving the dislocation by imposing an external force on the system, the dimer was manually moved, from one atomic position to the next, relative to the (pinned) dislocation.  The energy of this system was then minimized to produce the variation of energy (at zero applied stress) as a function of dislocation-dimer separation. To ensure that the dislocation does not move during energy minimization, 4 atoms localized at the core of one of the partial dislocations and far from the dimer are fixed (see \cite{dontsova14} for a detailed discussion).
Ideally, this interaction energy (denoted as $E_{interaction}$ here) should be given by the sum of the parelastic and dielastic contributions described above. Deriving an interaction force from a plot of E$_{interaction}$ versus distance is challenging as it requires the differentiation of a discontinuous set of discrete energy versus position points.  Rather than fit a smooth curve through these data points, and risk large errors in estimated slopes, and therefore forces, we will use only energies for quantitative comparison in these calculations.
%In order to avoid the ambiguity of deriving the interaction force between dislocations and dimer from the inherently discrete $E_{interaction}$-distance curve, only energies (and not forces) are compared here.    

In order to make a direct comparison between $E_{interaction}$ and the energies reported in Figs. \ref{fig:dimerprofile5} and \ref{fig:dimerprofile4} one needs to add the macroscopic elastic strain energy to $E_{interaction}$.  The light green squares in Figs. \ref{fig:dimerprofile5} and \ref{fig:dimerprofile4} show the results of adding the elastic strain energy to $E_{interaction}$, where the elastic strain energy was calculated based on the macroscopic elastic strain (total imposed strain minus the plastic strain) from the results obtained from loading.  

%As noted in the introduction, there have been several mechanisms proposed to explain the strength of solute clusters.  Here we attempt to unravel the energy-distance/force-distance curves for dimers I and II using two mechanisms; the elastic interaction between the solute atoms and dislocations and a `chemical' interaction related to the change in energy associated with the change in configuration between atoms as the dislocation passes.  As will be shown below, this second energy change is related to, but distinct from, the `order strengthening' contribution proposed by Starkink \emph{et al.}~\cite{SW09,SCR12}.  
%As has been noted before these two terms can be related to the classic para-elastic and di-elastic contributions in the classic solid solution strengthening literature \cite{}.

To estimate the parelastic interaction arising from the interaction between the stress field of the dislocations and the dilation of the solute atom we follow the classic approach of Cockhardt \emph{et al.} \cite{CSW55}.  The binding elastic energy between solute atoms acting as a point source of dilation of a dislocation is computed as,

\begin{equation}
E_{elastic} = -\delta\Omega \sum_{i=1}^2{\sigma_H^{\perp}\left(x_i,y_i\right)}
\label{eqn:para}
\end{equation}

\noindent where the sum is taken over the two solute atoms comprising the dimer.  The hydrostatic stresses at the location of the solute atoms ($\sigma_H^{\perp}\left(x_i,y_i\right)= \frac{1}{3}\sum_{ii} \sigma_{ii}$) have been evaluated from the per-atom virial stresses computed directly from a molecular statics simulation on a box containing a dislocation in pure Al.  In this way, the effects of finite simulation box, and the attendant image stresses, are automatically accounted for.
When the solute atoms are located in the stacking fault, the calculation of the elastic contribution to the energy is complicated. In this case, the local environment is not FCC Al and thus the relaxation volume ($\delta\Omega$) is not expected to be the same as that defined in Eqn. \ref{eqn:para}. For simplicity, however, we disregard this difference here, using for all calculations the value of $\delta\Omega$ defined for a pure homogeneous Al environment.

The value of the relaxation volume  ($\delta\Omega$)  can be obtained directly from atomistic calculations following the method outlined by Clouet et al. \cite{clouet18}, the value appropriate for Mg in Al using the current interatomic potential is given in Table \ref{table:elastic_constants}.  

The chemical (or dielastic \cite{F61,F63}) contribution to the interaction energy profile between the dislocation and dimer has been evaluated following the method outlined by Ma \emph{et al.} \cite{MFPRN15} based on the method originally developed by Yasi \emph{et al.} \cite{YHT10}.   This contribution arises from the change in generalized stacking fault energy (or $\gamma$-surface) resulting from the presence of the solute dimer and changes in its configuration as a dislocation passes.   This contribution, denoted as $E_{slip}$ following \cite{MFPRN15}, can be calculated as the difference in the generalized stacking fault curve obtained for pure aluminum and, separately, for the same system containing a dimer.  Thus,

\begin{equation}
E_{slip} = \gamma_{dimer}\left(x_p,y_p\right) - \gamma_{Al}\left(x_p,y_p\right)
\label{eqn:eslip}
\end{equation}

\noindent where $\left( x_p, y_p \right)$ represents the path in the slip plane taken by atoms as the dislocation passes and $\gamma_{Al}$ and $\gamma_{dimer}$ represent the generalized stacking fault curves for pure Al and for the same simulation box but containing the dimer of interest.   

%The two model contributions considered above are for the case where the interaction between dislocation and solute atoms occurs with no imposed load.  This is clearly different from the conditions corresponding to Figs. \ref{fig:dimerprofile5} and \ref{fig:dimerprofile4} where the results are obtained under a finite (varying) load.  To make a more direct comparision between the $E_{elastic}$ and $E_{slip}$ a different series of molecular statics simulations were performed in a simulation box consisting of pure Al and a single (split) edge dislocation.  

%A solute dimer was introduced into the simulation box and its position was manually moved from one atomic position to the next along the glide direction of the dislocation.  At each position a molecular statics simulation was performed and the change in energy (relative to the energy of a system with dislocation and solute at infinite separation) was recorded.  In this way the interaction energy ($E_{interaction}$) could be obtained on a simulation box under no external applied stress.\commentAV{How did we make sure that the dislocation did not run away?}

\begin{figure}[htbp]
\centering     %%% not \center
\begin{tikzpicture}
\node [above] at (0,0) {\includegraphics[width=0.8\textwidth]{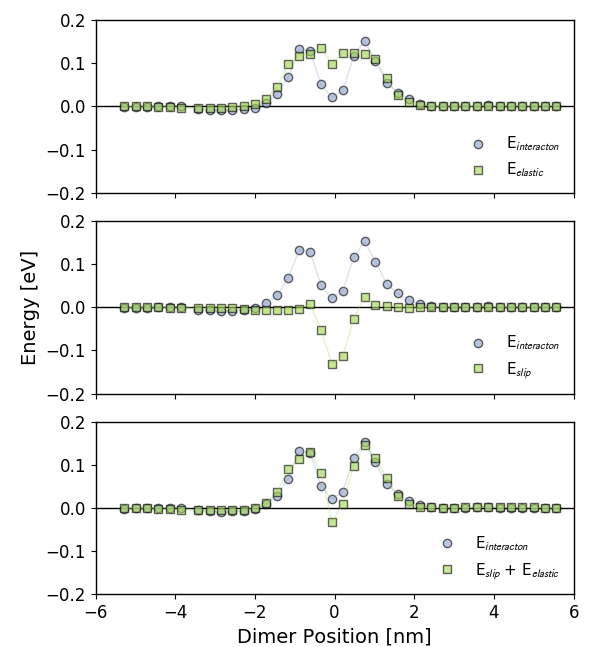}};
\node at (5,8.85) {(a)};
\node at (5,5.6) {(b)};
\node at (5,2.35) {(c)};
\end{tikzpicture}
\caption{Energy distance profiles for dislocation/dimer I interactions.  $E_{interaction}$ is the energy calculated directly from molecular statics simulations (no applied force) where the centre of the dislocation's stacking fault is located at a dimer position of `0'.  (a) shows the comparison between $E_{interaction}$ and the parelastic (Eqn. \ref{eqn:para}) contribution to the interaction energy,  (b) shows the comparison with $E_{slip}$ (Eqn. \ref{eqn:eslip}), and (c) compares the atomistically computed $E_{interaction}$ and $E_{elastic}+E_{slip}$.}
\label{fig:modeldimer5}
\end{figure}
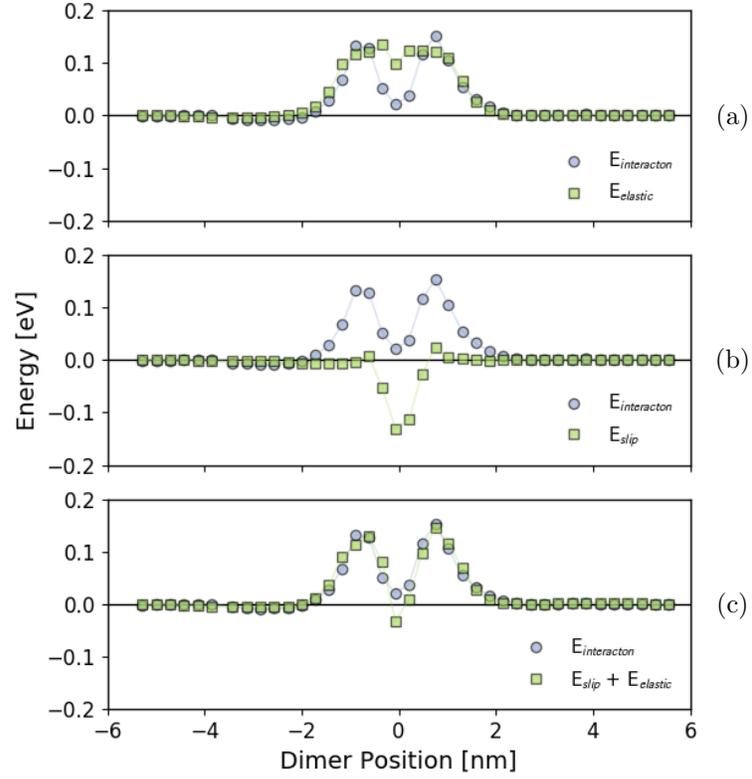

Fig. \ref{fig:modeldimer5} shows the comparison between $E_{interaction}$ and the two model contributions $E_{elastic}$ (Eqn. \ref{eqn:para}) and $E_{slip}$ (Eqn \ref{eqn:eslip}), in the case of the dimer with the highest strength (dimer I).  First, one can see that the shape of $E_{interaction}$ follows closely that presented in Fig. \ref{fig:dimerprofile5}, the additional energy in the later arising from the macroscopically imposed load on the simulation box.    Fig. \ref{fig:modeldimer5}(a) clearly shows that the rise in interaction energy that occurs as the dislocation first approaches the dimer comes predominantly from the elastic interaction between the dislocation and dimer.  Once the dimer crosses the leading partial, however, elasticity alone fails to predict the drop in energy associated with the dimer lying within the stacking fault.  This drop is captured by $E_{slip}$ (see Fig. \ref{fig:modeldimer5}(b)), and is associated with the change in configuration between the Mg atoms when they are sitting within the HCP environment of the stacking fault.  As the dimer passes the trailing dislocation, the energy is seen to drop again, following $E_{slip}$, with the energy returning to zero as there is no net change in the configuration of the dimer in this example.

One can see in Fig. \ref{fig:modeldimer5}(c) that, in this case, the combination of the two models ($E_{elastic} + E_{slip}$) captures all of the main features of the $E_{interaction}$-distance curve.  The maximum resistance provided by this dimer is seen from Fig. \ref{fig:dimerprofile5} to occur when the dimer passes the trailing dislocation, this only being slightly higher than the force required for the leading partial to pass the dimer.  The similarity of these forces is consistent with the similarity of the shape of the $E_{interaction}$-distance curve in Fig. \ref{fig:dimerprofile5} on the leading edge of the two peaks, recalling that the force is the slope of the energy-distance curve.  Observing the shape of the $E_{elastic}$ and $E_{slip}$ curves shows that the maximum force is dominated by the $E_{slip}$ contribution \emph{even though no net change in the configuration of the solute atoms occurs after the dislocation has passed}.  

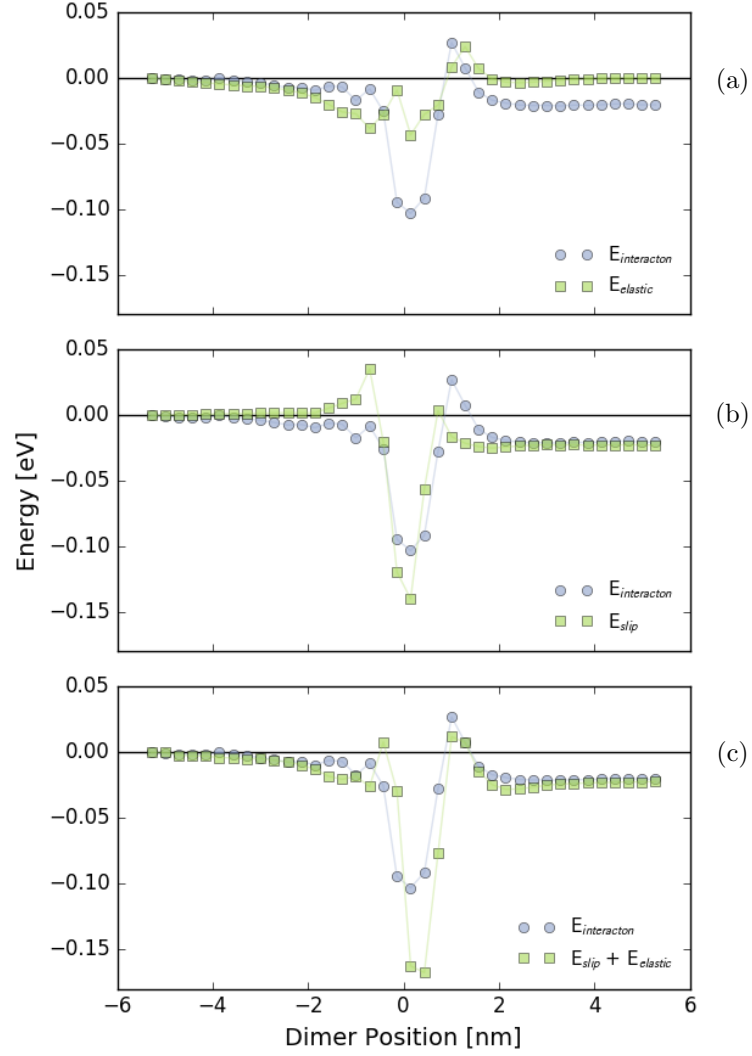
\begin{figure}[htbp]
\centering     %%% not \center
\begin{tikzpicture}
\node [above] at (0,0) {\includegraphics[width=0.8\textwidth]{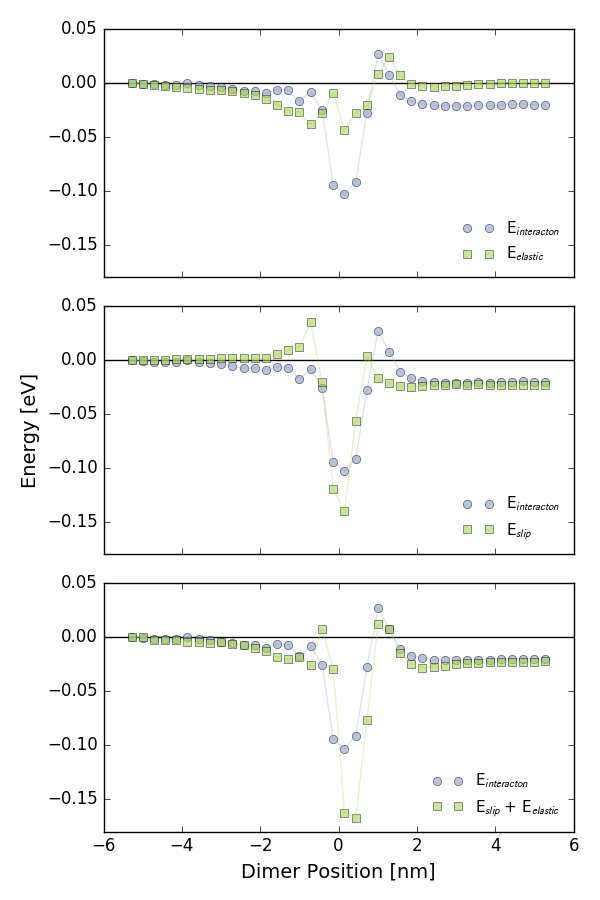}};
\node at (5,13.3) {(a)};
\node at (5,8.85) {(b)};
\node at (5,4.35) {(c)};
\end{tikzpicture}
\caption{Energy distance profiles for dislocation/dimer II interactions.  $E_{interaction}$ is the energy calculated directly from molecular statics simulations (no applied force) where the centre of the dislocation's stacking fault is located at a dimer position of `0'.  (a) shows the comparison between $E_{interaction}$ and the paraelastic (Eqn. \ref{eqn:para}) contribution to the interaction energy,  (b) shows the comparison with $E_{slip}$ (Eqn. \ref{eqn:eslip}), and (c) compares the atomistically computed $E_{interaction}$ and $E_{elastic}+E_{slip}$. }
\label{fig:modeldimer4}
\end{figure}

Turning to the dimer with the second highest strength (\textit{i.e.} dimer II), Fig. \ref{fig:modeldimer4}, we can see that the $E_{elastic}$ provides a relatively small contribution to the overall energy-distance profile.  This is not unexpected given the fact that the dimer straddles the glide plane.  In this example, the solute atoms comprising the dimer (each having the same relaxation volume) sit on both sides of the glide plane (one sits on the tensile side, and the other on the compression side). In terms of the total $E_{elastic}$ then, the contributions from both solutes nearly cancel out one another. They do not completely as they are slightly offset from one another along the glide direction.  This matches well with the observed small change in $E_{elastic}$ (see Fig. \ref{fig:modeldimer4}(a)) as the leading partial approaches the dimer and the trailing partial moves away from it.  

While $E_{elastic}$ is small in this instance, there is a large drop in $E_{slip}$ associated with the dimer entering the stacking fault (see Fig. \ref{fig:modeldimer4}(b)).  In this case, the sum of $E_{slip} + E_{elastic}$ overestimates the energy associated with the dimer located in the stacking fault (see Fig. \ref{fig:modeldimer4}(c)).  A large part of this error can be attributed to the fact that we use a relaxation volume for the Mg atoms (cf. Eqn. \ref{eqn:para}) for a Mg atom located in a perfect FCC Al lattice, not in a stacking fault.   After the dislocation fully bypasses the dimer, a net change in $E_{interaction}$ is seen.  This is reflected in the net change in $E_{slip}$ and is directly attributable to the change in configuration between Mg atoms following the passage of the dislocation.  

As in the case of Fig. \ref{fig:dimerprofile5}, the configuration corresponding to the largest force in Fig. \ref{fig:dimerprofile4} corresponds to the dimer passing by the trailing dislocation (leading to the dimer exiting the stacking fault).  This is more readily apparent from Fig. \ref{fig:modeldimer4}, where the slope of the $E_{interaction}$-distance curve appears highest for the portion of the curve corresponding to the dimer exiting the stacking fault, the rapid change in $E_{interaction}$ in this portion of the curve being dominated by $E_{slip}$.

The primary conclusion from the results shown in Figs. \ref{fig:modeldimer5} and \ref{fig:modeldimer4} is that $E_{slip}$ is the main contributor to the strength of the two strongest dimers.  This also helps to explain why the strengthening observed in Al-Mg alloys is as small as it is, despite the fact that Mg is a strong solid solution strengthener in Al \cite{P95}.
%Both dimers considered here have Mg atoms as nearest neighbours which is actually energetically unfavourable for Mg atoms, the second nearest neighbour positions being energetically favourable consistent with experimental observations~\cite{SKT82} and prior reports related to this interatomic potential~\cite{ZP04,dontsova14}.  
%This is seen here in the fact that the energy is \emph{reduced} when the dislocation passes by dimer II (Fig. \ref{fig:modeldimer4}).  The passage of the dislocation in this example leads to a change from first to second nearest neighbour configuration for the Mg atoms and therefore a reduction in energy.  

\section{Consequences for the prediction of cluster strengthening}

%\subsection{Re-visiting the Starink Order Strengthening Model}

A key conclusion arising from the results in Figs. \ref{fig:dimerprofile5} and \ref{fig:dimerprofile4} is that the maximum force for both of the strongest dimers is achieved when the trailing dislocation overcomes the dimer and that this force is dominated by the slope of the $E_{slip}$-distance curve.  The lack of contribution from $E_{elastic}$ when the dimer resides within the stacking fault is due to the fact that the hydrostatic stress near the stacking fault and between the two partial dislocations is nearly constant.  This is reflected in the fact that the value of $E_{elastic}$ is nearly constant when the dimer is situated between the two partials (Figs.  \ref{fig:modeldimer5} and \ref{fig:modeldimer4}).  The small variation seen in Fig. \ref{fig:modeldimer4} arises when the dimer straddles the leading partial. The variation in $E_{slip}$ as a source of dimer strength represents a generalization of the  `order strengthening' model proposed by Starink \emph{et al.} ~\cite{SW09,SCR12}.  In the original `order strengthening' model, it was the change in energy caused by the permanent configurational change between atoms comprising a dimer (or cluster) due to the passage of a perfect edge dislocation.  The force to make this change was then estimated as $\Delta H/b$ where $\Delta H$ was envisioned as the energy required to make the configurational change and $b$ is the magnitude of the Burgers vector of the dislocation.  The present results show that a high resistance to dislocation motion can be obtained from a similar source in the presence of a split edge dislocation (similar results arise from a split screw dislocation) even if no net configuration change occurs between the atoms comprising the dimer since the passage of the dimer across both partials and the stacking fault leads to configurational changes.  Moreover, the correct distance to estimate the slope of the energy-distance curve is not the Burgers vector of the perfect dislocation, but rather a distance better approximated as half the width of the stacking fault.  

Another implicit assumption in the Starink `order strengthening' model is that the energy-distance curve is linear with $\Delta H/b$ being a good approximation to the force.  Evaluating the energy-distance (and force-distance) profiles from a number of dimers (including those shown shown in Figs. \ref{fig:dimerprofile5} and \ref{fig:dimerprofile4}) show that the energy-distance profiles between the energy minima corresponding to the dimer located in the middle of the stacking fault and the energy maxima just prior to the dislocation escaping the stacking fault are better described as parabolic, the force-distance curves being linear. The challenge, given that $E_{interaction}$ varies linearly with distance is that it does not give a precise value of the force required for the dislocation to break free of the dimer, as shown in Figs \ref{fig:modeldimer5} and \ref{fig:modeldimer4}. Owing to this, a simple model for dimer strength based on a variation of the Starink model is difficult.

%\subsection{Is Eslip always the dominant factor in dimer strength?}

The strength of the two strongest dimers here is dominated by the dielastic contribution to the interaction energy. However, it cannot be extrapolated from these results that this would always be the case. As an illustration, consider a hypothetical version of dimer I having the same value of $E_{slip}$ but a larger $\delta \Omega$. For this hypothetical dimer, the strength would be dominated by the entry of the dimer into the stacking fault.  However, changing the volume difference would not have drastically changed the result for dimer II: although the strength of the initial interaction would have changed, the critical configuration would have remained the same. In the case of dimers comprised of the same alloying element, the elastic interaction for those straddeling the glide plane would be always small owing to the cancelling of the effects from the opposite interactions with the tensile and compressive side of the two dislocations. In the case of co-clusters, the effect can be opposite with dimers comprised of solute on the same side of the glide plane exhibiting an elastic interaction that nearly cancels out if their volume difference with Al atoms is opposite.  In either case, the elastic interaction will only be effective for some fraction of the dimers.  In the present case only 2 out of 14 cases would be effective.  In the case of co-clusters, one would imagine this ratio to be reversed, thus having the potential for a strong elastic interaction; though this will be strongest for dimers where the two solute atoms sit nearly directly on top of one another (e.g. dimers XI and VIII).  

%\subsection{How do the above results translate to clusters of larger size?}

The results presented in this paper for the strength of dimer are difficult to extrapolate to clusters with higher number of solute atoms as increasing the number of solute atoms comprising the cluster increases the number of configurations possible exponentially. However, we can suppose that as a cluster becomes larger, the range of interaction for the parelastic and dielastic contributions can increase. In the case of dimers, the parelastic contribution ($E_{elastic}$) to the force is limited to configurations where the dimer is located outside of the stacking fault whereas the dielastic contribution ($E_{slip}$) is limited to cases where it is located within the stacking fault. As the cluster grows, and if its diameter is larger than the distance separating partial dislocations, configurations should exist where a part of the dimer will be located outside of the stacking fault and a part will be located inside.  
In order for both parelastic and dielastic contributions to contribute additively, some portion of the cluster needs to be located inside the stacking fault (for a dielastic contribution) while another portion needs to be outside of the stacking fault (for a large parelastic contribution).  This suggests the largest contribution would occur when the cluster is centered on one of the partial dislocations.
%In this case, it is possible for both contributions to be additive, and thus for both to be important. 
This is distinct for the case of dimers where the distinct separation of the regions of interaction mean that the only place where the two can have some overlap is at the position where a dimer is located at one of the partials.

%Finally, as the strongest configurations for dimers seen here are not energetically favourable for Mg atoms~\commentAV{cite: Literature on the precipitation of Al3Mg}, it can be seen why the Al-Mg system presents weak cluster hardening while Al-Mg-Si and Al-Mg-Cu are strong cluster hardening systems.

\section{Conclusion}

Using a `model' Al-Mg alloy, we have been able to evaluate the contributions of classic `continuum' chemical and elastic effects to the glide resistance of solute dimers.  It was shown that the concept of `order strengthening' proposed by Starink \emph{et al.} can be generalized to consider the variations in energy associated with changes in $\gamma$-surface leading to the conclusion that strong dimers do not need to be ones whose configuration is changed after the passage of a dislocation.  It was also shown that the classic approach of Cockhardt \emph{et al.} \cite{CSW55} adequately allows the variation of the parelastic contribution ($E_{elastic}$) to be predicted when each of the solute atoms is separately considered as a point source of dilation.  Comparison of these quasi-analytical calculations to the results from molecular statics simulations reveals that both mechanisms can lead to dimers with high strength but that the dielastic contribution ($E_{slip}$) has the potential to impact on more dimer configurations given that it is less sensitive to their configuration relative to the dislocation.  Of course, these results only relate to the strength of individual dimers. The strength of a cluster strengthened alloy needs to further consider the statistical distribution of these solute clusters within the alloy.

\section*{Declaration of Interests}
The authors declare that they have no known competing financial interests or personal relationships that could have appeared to influence the work reported in this paper.

\section*{Acknowledgement}

The authors would like to gratefully acknowledge the financial support from Natural Sciences and Engineering Research Council of Canada (NSERC) through the NSERC Discovery Grant program.

%\printbibliography
%\bibliographystyle{ieee}
\bibliographystyle{elsarticle-num}
\bibliography{literature}

\begin{thebibliography}{10}
\expandafter\ifx\csname url\endcsname\relax
  \def\url#1{\texttt{#1}}\fi
\expandafter\ifx\csname urlprefix\endcsname\relax\def\urlprefix{URL }\fi
\expandafter\ifx\csname href\endcsname\relax
  \def\href#1#2{#2} \def\path#1{#1}\fi

\bibitem{RSP97}
S.~P. Ringer, T.~Sakurai, I.~J. Polmear, Origins of hardening in aged
  {Al--Gu--Mg--(Ag)} alloys, Acta Materialia 45~(9) (1997) 3731 -- 3744.
\newblock \href {http://dx.doi.org/10.1016/S1359-6454(97)00039-6}
  {\path{doi:10.1016/S1359-6454(97)00039-6}}.

\bibitem{P95}
I.~J. Polmear, Light alloys : metallurgy of the light metals, Arnold, London,
  1995.

\bibitem{RH00}
S.~P. Ringer, K.~Hono,
  \href{http://www.sciencedirect.com/science/article/pii/S1044580399000510}{Microstructural
  evolution and age hardening in aluminium alloys: Atom probe field{--}ion
  microscopy and transmission electron microscopy studies}, Materials
  Characterization 44~(1–2) (2000) 101 -- 131.
\newblock \href
  {http://dx.doi.org/http://dx.doi.org/10.1016/S1044-5803(99)00051-0}
  {\path{doi:http://dx.doi.org/10.1016/S1044-5803(99)00051-0}}.
\newline\urlprefix\url{http://www.sciencedirect.com/science/article/pii/S1044580399000510}

\bibitem{MVRSP13}
R.~Marceau, A.~de~Vaucorbeil, G.~Sha, S.~Ringer, W.~Poole,
  \href{http://www.sciencedirect.com/science/article/pii/S1359645413006290}{Analysis
  of strengthening in {AA6111} during the early stages of aging: Atom probe
  tomography and yield stress modelling}, Acta Materialia 61~(19) (2013) 7285
  -- 7303.
\newblock \href
  {http://dx.doi.org/http://dx.doi.org/10.1016/j.actamat.2013.08.033}
  {\path{doi:http://dx.doi.org/10.1016/j.actamat.2013.08.033}}.
\newline\urlprefix\url{http://www.sciencedirect.com/science/article/pii/S1359645413006290}

\bibitem{SKT82}
T.~Sato, Y.~Kojima, T.~Takahashi,
  \href{http://dx.doi.org/10.1007/BF02642874}{Modulated structures and {GP}
  zones in {Al--Mg} alloys}, Metallurgical Transactions A 13~(8) (1982)
  1373--1378.
\newblock \href {http://dx.doi.org/10.1007/BF02642874}
  {\path{doi:10.1007/BF02642874}}.
\newline\urlprefix\url{http://dx.doi.org/10.1007/BF02642874}

\bibitem{OO84}
K.~Osamura, T.~Ogura, \href{http://dx.doi.org/10.1007/BF02644557}{Metastable
  phases in the early stage of precipitation in {Al-Mg} alloys}, Metallurgical
  Transactions A 15~(5) (1984) 835--842.
\newblock \href {http://dx.doi.org/10.1007/BF02644557}
  {\path{doi:10.1007/BF02644557}}.
\newline\urlprefix\url{http://dx.doi.org/10.1007/BF02644557}

\bibitem{KN63}
A.~Kelly, R.~B. Nicholson, Precipitation hardening, Progress in Materials
  Science 10 (1963) 151--391.

\bibitem{MSFDR10}
R.~Marceau, G.~Sha, R.~Ferragut, A.~Dupasquier, S.~Ringer,
  \href{http://www.sciencedirect.com/science/article/pii/S1359645410002958}{Solute
  clustering in {Al–Cu–Mg} alloys during the early stages of elevated
  temperature ageing}, Acta Materialia 58~(15) (2010) 4923 -- 4939.
\newblock \href
  {http://dx.doi.org/http://dx.doi.org/10.1016/j.actamat.2010.05.020}
  {\path{doi:http://dx.doi.org/10.1016/j.actamat.2010.05.020}}.
\newline\urlprefix\url{http://www.sciencedirect.com/science/article/pii/S1359645410002958}

\bibitem{SCR12}
M.~J. Starink, L.~F. Cao, P.~A. Rometsch, A model for the thermodynamics of and
  strengthening due to co--clusters in {Al--Mg--Si-based} alloys, Acta
  Materialia 60~(10) (2012) 4194 -- 4207.
\newblock \href {http://dx.doi.org/10.1016/j.actamat.2012.04.032}
  {\path{doi:10.1016/j.actamat.2012.04.032}}.

\bibitem{ZHLSL15}
Q.~Zhao, B.~Holmedal, Y.~Li, E.~Sagvolden, O.~M. Løvvik,
  \href{http://www.sciencedirect.com/science/article/pii/S0921509314014841}{Multi{-}component
  solid solution and cluster hardening of {Al–Mn–Si} alloys}, Materials
  Science and Engineering: A 625 (2015) 153 -- 157.
\newblock \href
  {http://dx.doi.org/http://dx.doi.org/10.1016/j.msea.2014.12.006}
  {\path{doi:http://dx.doi.org/10.1016/j.msea.2014.12.006}}.
\newline\urlprefix\url{http://www.sciencedirect.com/science/article/pii/S0921509314014841}

\bibitem{Z14}
Q.~Zhao,
  \href{http://www.sciencedirect.com/science/article/pii/S1359646214001638}{Cluster
  strengthening in aluminium alloys}, Scripta Materialia 84–85 (2014) 43 --
  46.
\newblock \href
  {http://dx.doi.org/http://dx.doi.org/10.1016/j.scriptamat.2014.04.018}
  {\path{doi:http://dx.doi.org/10.1016/j.scriptamat.2014.04.018}}.
\newline\urlprefix\url{http://www.sciencedirect.com/science/article/pii/S1359646214001638}

\bibitem{SW09}
M.~Starink, S.~Wang,
  \href{http://www.sciencedirect.com/science/article/pii/S1359645409000482}{The
  thermodynamics of and strengthening due to co-clusters: {G}eneral theory and
  application to the case of {Al–Cu–Mg} alloys}, Acta Materialia 57~(8)
  (2009) 2376 -- 2389.
\newblock \href
  {http://dx.doi.org/http://dx.doi.org/10.1016/j.actamat.2009.01.021}
  {\path{doi:http://dx.doi.org/10.1016/j.actamat.2009.01.021}}.
\newline\urlprefix\url{http://www.sciencedirect.com/science/article/pii/S1359645409000482}

\bibitem{PRBM06}
L.~Proville, D.~Rodney, Y.~Br\'echet, G.~Martin,
  \href{http://www.tandfonline.com/doi/abs/10.1080/14786430600567721}{Atomic--scale
  study of dislocation glide in a model solid solution}, Philosophical Magazine
  86~(25-26) (2006) 3893--3920.
\newblock \href {http://dx.doi.org/10.1080/14786430600567721}
  {\path{doi:10.1080/14786430600567721}}.
\newline\urlprefix\url{http://www.tandfonline.com/doi/abs/10.1080/14786430600567721}

\bibitem{PP11}
S.~Patinet, L.~Proville,
  \href{http://www.tandfonline.com/doi/abs/10.1080/14786435.2010.543649}{Dislocation
  pinning by substitutional impurities in an atomic--scale model for {Al(Mg)}
  solid solutions}, Philosophical Magazine 91~(11) (2011) 1581--1606.
\newblock \href {http://dx.doi.org/10.1080/14786435.2010.543649}
  {\path{doi:10.1080/14786435.2010.543649}}.
\newline\urlprefix\url{http://www.tandfonline.com/doi/abs/10.1080/14786435.2010.543649}

\bibitem{XP06}
Z.~Xu, R.~C. Picu,
  \href{http://stacks.iop.org/0965-0393/14/i=2/a=005}{Dislocation--solute
  cluster interaction in {Al--Mg} binary alloys}, Modelling and Simulation in
  Materials Science and Engineering 14~(2) (2006) 195.
\newline\urlprefix\url{http://stacks.iop.org/0965-0393/14/i=2/a=005}

\bibitem{DRS14}
E.~Dontsova, J.~Rottler, C.~W. Sinclair,
  \href{http://link.aps.org/doi/10.1103/PhysRevB.90.174102}{Solute-defect
  interactions in {Al-Mg} alloys from diffusive variational gaussian
  calculations}, Phys. Rev. B 90 (2014) 174102.
\newblock \href {http://dx.doi.org/10.1103/PhysRevB.90.174102}
  {\path{doi:10.1103/PhysRevB.90.174102}}.
\newline\urlprefix\url{http://link.aps.org/doi/10.1103/PhysRevB.90.174102}

\bibitem{P09}
S.~Patinet, Durcissement par solution solide dans les alliages m\'etalliques
  {CFC}, Ph.D. thesis, Universit\'e Paris-Sud XI (2009).

\bibitem{vaucorbeil15}
A.~de~Vaucorbeil,
  \href{https://open.library.ubc.ca/cIRcle/collections/24/items/1.0221372}{On
  the origin of cluster strengthening in aluminum alloys}, Ph.D. thesis,
  University of British Columbia (2015).
\newblock \href {http://dx.doi.org/http://dx.doi.org/10.14288/1.0221372}
  {\path{doi:http://dx.doi.org/10.14288/1.0221372}}.
\newline\urlprefix\url{https://open.library.ubc.ca/cIRcle/collections/24/items/1.0221372}

\bibitem{BOR09}
D.~J. Bacon, Y.~N. Osetsky, D.~Rodney, Dislocation--obstacle interactions at
  the atomic level, in: J.~P. Hirth, L.~Kubin (Eds.), Dislocations in Solids,
  Elsevier, 2009.

\bibitem{Rodney07}
D.~Rodney,
  \href{https://link.aps.org/doi/10.1103/PhysRevB.76.144108}{Activation
  enthalpy for kink-pair nucleation on dislocations: Comparison between static
  and dynamic atomic-scale simulations}, Phys. Rev. B 76 (2007) 144108.
\newblock \href {http://dx.doi.org/10.1103/PhysRevB.76.144108}
  {\path{doi:10.1103/PhysRevB.76.144108}}.
\newline\urlprefix\url{https://link.aps.org/doi/10.1103/PhysRevB.76.144108}

\bibitem{LA98}
X.-Y. Liu, J.~Adams,
  \href{http://www.sciencedirect.com/science/article/pii/S135964549800038X}{Grain--boundary
  segregation in {Al--10\%Mg} alloys at hot working temperatures}, Acta
  Materialia 46~(10) (1998) 3467 -- 3476.
\newblock \href
  {http://dx.doi.org/http://dx.doi.org/10.1016/S1359-6454(98)00038-X}
  {\path{doi:http://dx.doi.org/10.1016/S1359-6454(98)00038-X}}.
\newline\urlprefix\url{http://www.sciencedirect.com/science/article/pii/S135964549800038X}

\bibitem{OHC06}
D.~L. Olmsted, L.~G.~H. Jr., W.~A. Curtin,
  \href{http://www.sciencedirect.com/science/article/pii/S0022509606000354}{Molecular
  dynamics study of solute strengthening in {Al/Mg} alloys}, Journal of the
  Mechanics and Physics of Solids 54~(8) (2006) 1763 -- 1788.
\newblock \href
  {http://dx.doi.org/http://dx.doi.org/10.1016/j.jmps.2005.12.008}
  {\path{doi:http://dx.doi.org/10.1016/j.jmps.2005.12.008}}.
\newline\urlprefix\url{http://www.sciencedirect.com/science/article/pii/S0022509606000354}

\bibitem{ZP04}
D.~Zhang, R.~C. Picu,
  \href{http://stacks.iop.org/0965-0393/12/i=1/a=011}{Solute clustering in
  {Al--Mg} binary alloys}, Modelling and Simulation in Materials Science and
  Engineering 12~(1) (2004) 121.
\newline\urlprefix\url{http://stacks.iop.org/0965-0393/12/i=1/a=011}

\bibitem{dontsova14}
E.~Dontsova, J.~Rottler, C.~W. Sinclair,
  \href{https://link.aps.org/doi/10.1103/PhysRevB.90.174102}{Solute-defect
  interactions in al-mg alloys from diffusive variational gaussian
  calculations}, Phys. Rev. B 90 (2014) 174102.
\newblock \href {http://dx.doi.org/10.1103/PhysRevB.90.174102}
  {\path{doi:10.1103/PhysRevB.90.174102}}.
\newline\urlprefix\url{https://link.aps.org/doi/10.1103/PhysRevB.90.174102}

\bibitem{lammps_main}
{LAMMPS} molecular dynamics simulator, \url{http://lammps.sandia.gov/}.

\bibitem{ovito}
A.~Stukowski,
  \href{http://stacks.iop.org/0965-0393/18/i=1/a=015012}{Visualization and
  analysis of atomistic simulation data with ovito–the open visualization
  tool}, Modelling and Simulation in Materials Science and Engineering 18~(1)
  (2010) 015012.
\newline\urlprefix\url{http://stacks.iop.org/0965-0393/18/i=1/a=015012}

\bibitem{VMSB64}
J.~Vallin, M.~Mongy, K.~Salama, O.~Beckman,
  \href{http://scitation.aip.org/content/aip/journal/jap/35/6/10.1063/1.1713749}{Elastic
  constants of aluminum}, Journal of Applied Physics 35~(6) (1964) 1825--1826.
\newblock \href {http://dx.doi.org/http://dx.doi.org/10.1063/1.1713749}
  {\path{doi:http://dx.doi.org/10.1063/1.1713749}}.
\newline\urlprefix\url{http://scitation.aip.org/content/aip/journal/jap/35/6/10.1063/1.1713749}

\bibitem{WP71}
K.~H. Westmacott, R.~L. Peck,
  \href{http://dx.doi.org/10.1080/14786437108216407}{A rationalization of
  secondary defect structures in aluminium{--}based alloys}, Philosophical
  Magazine 23~(183) (1971) 611--622.
\newblock \href {http://dx.doi.org/10.1080/14786437108216407}
  {\path{doi:10.1080/14786437108216407}}.
\newline\urlprefix\url{http://dx.doi.org/10.1080/14786437108216407}

\bibitem{R82}
R.~H. Rautioaho, \href{http://dx.doi.org/10.1002/pssb.2221120108}{An
  interatomic pair potential for aluminium calculation of stacking fault
  energy}, physica status solidi (b) 112~(1) (1982) 83--89.
\newblock \href {http://dx.doi.org/10.1002/pssb.2221120108}
  {\path{doi:10.1002/pssb.2221120108}}.
\newline\urlprefix\url{http://dx.doi.org/10.1002/pssb.2221120108}

\bibitem{H96}
P.~HAASEN, Physical Metallurgy, third edition Edition, Cambridge University
  Press, 1996.

\bibitem{CSW55}
A.~Cochardt, G.~Schoek, H.~Wiedersich,
  \href{http://www.sciencedirect.com/science/article/pii/0001616055901115}{Interaction
  between dislocations and interstitial atoms in body--centered cubic metals},
  Acta Metallurgica 3~(6) (1955) 533 -- 537.
\newblock \href
  {http://dx.doi.org/http://dx.doi.org/10.1016/0001-6160(55)90111-5}
  {\path{doi:http://dx.doi.org/10.1016/0001-6160(55)90111-5}}.
\newline\urlprefix\url{http://www.sciencedirect.com/science/article/pii/0001616055901115}

\bibitem{clouet18}
E.~Clouet, C.~Varvenne, T.~Jourdan,
  \href{https://doi.org/10.1016/j.commatsci.2018.01.053}{Elastic modeling of
  point-defects and their interaction}, Computational Materials Science 147
  (2018) 49--63.
\newblock \href {http://dx.doi.org/10.1016/j.commatsci.2018.01.053}
  {\path{doi:10.1016/j.commatsci.2018.01.053}}.
\newline\urlprefix\url{https://doi.org/10.1016/j.commatsci.2018.01.053}

\bibitem{F61}
R.~L. Fleischer,
  \href{http://www.sciencedirect.com/science/article/pii/0001616061902425}{Solution
  hardening}, Acta Metallurgica 9~(11) (1961) 996 -- 1000.
\newblock \href
  {http://dx.doi.org/http://dx.doi.org/10.1016/0001-6160(61)90242-5}
  {\path{doi:http://dx.doi.org/10.1016/0001-6160(61)90242-5}}.
\newline\urlprefix\url{http://www.sciencedirect.com/science/article/pii/0001616061902425}

\bibitem{F63}
R.~L. Fleischer,
  \href{http://www.sciencedirect.com/science/article/pii/000161606390213X}{Substitutional
  solution hardening}, Acta Metallurgica 11~(3) (1963) 203 -- 209.
\newblock \href
  {http://dx.doi.org/http://dx.doi.org/10.1016/0001-6160(63)90213-X}
  {\path{doi:http://dx.doi.org/10.1016/0001-6160(63)90213-X}}.
\newline\urlprefix\url{http://www.sciencedirect.com/science/article/pii/000161606390213X}

\bibitem{MFPRN15}
D.~Ma, M.~Fri\'ak, J.~von Pezold, D.~Raabe, J.~Neugebauer,
  \href{http://www.sciencedirect.com/science/article/pii/S1359645414008088}{Computationally
  efficient and quantitatively accurate multiscale simulation of
  solid{--}solution strengthening by ab initio calculation}, Acta Materialia 85
  (2015) 53 -- 66.
\newblock \href
  {http://dx.doi.org/http://dx.doi.org/10.1016/j.actamat.2014.10.044}
  {\path{doi:http://dx.doi.org/10.1016/j.actamat.2014.10.044}}.
\newline\urlprefix\url{http://www.sciencedirect.com/science/article/pii/S1359645414008088}

\bibitem{YHT10}
J.~A. Yasi, L.~G.~H. Jr., D.~R. Trinkle,
  \href{http://www.sciencedirect.com/science/article/pii/S1359645410004106}{Firs{t-p}rinciples
  data for solid-solution strengthening of magnesium: From geometry and
  chemistry to properties}, Acta Materialia 58~(17) (2010) 5704 -- 5713.
\newblock \href
  {http://dx.doi.org/http://dx.doi.org/10.1016/j.actamat.2010.06.045}
  {\path{doi:http://dx.doi.org/10.1016/j.actamat.2010.06.045}}.
\newline\urlprefix\url{http://www.sciencedirect.com/science/article/pii/S1359645410004106}

\end{thebibliography}

\end{document}